\documentclass[12pt,preprint]{aastex}

\usepackage{graphicx}
\usepackage{stmaryrd}
\usepackage{multirow}
\usepackage{placeins}

\usepackage{natbib}
\newcommand{\simi}{\ensuremath{\sim}}
\newcommand{\elec}{e\ensuremath{^{-}}}

\newcommand{\st}{\ensuremath{^{\textrm{\scriptsize st}}} }
\newcommand{\nd}{\ensuremath{^{\textrm{\scriptsize nd}}} }
\newcommand{\rd}{\ensuremath{^{\textrm{\scriptsize rd}}} }
\newcommand{\tth}{\ensuremath{^{\textrm{\scriptsize th}}} }
\newcommand{\ste}{\ensuremath{^{\textrm{\scriptsize st}}}}
\newcommand{\nde}{\ensuremath{^{\textrm{\scriptsize nd}}}}

\newcommand{\modif}{}

\shorttitle{Exoplanet XO-2\MakeLowercase{b}}
\shortauthors{Crouzet et al.}

\begin{document}


\title{Transmission spectroscopy of exoplanet XO-2\MakeLowercase{b} observed with HST NICMOS}


\author{N. Crouzet\altaffilmark{1}, P. R. McCullough\altaffilmark{1}, C. Burke\altaffilmark{2}, D. Long\altaffilmark{1}}

\affil{\altaffilmark{1} Space Telescope Science Institute, 3700 San Martin Drive, Baltimore, MD 21218, USA}
\affil{\altaffilmark{2} NASA Ames Research Center, Moffett Field, CA, 94035, USA}

\email{crouzet@stsci.edu}

%




\begin{abstract}
Spectroscopy during planetary transits is a powerful tool to probe exoplanet atmospheres. We present the near-infrared transit spectroscopy of XO-2b obtained with HST NICMOS. Uniquely for NICMOS transit spectroscopy, a companion star of similar properties to XO-2 is present in the field of view. We derive improved star and planet parameters through a photometric white-light analysis. We show a clear correlation of the spectrum noise with instrumental parameters, in particular the angle of the spectral trace on the detector. An MCMC method using a decorrelation from instrumental parameters is used to extract the planetary spectrum. Spectra derived independently from each of the 3 visits have a RMS of 430, 510, and 1000~ppm respectively. The same analysis is performed on the companion star after numerical injection of a transit with a depth constant at all wavelengths. The extracted spectra exhibit residuals of similar amplitude as for XO-2, which represent the level of remaining NICMOS systematics. This shows that extracting planetary spectra is at the limit of NICMOS' capability. We derive a spectrum for the planet XO-2b using the companion star as a reference. \modif{The derived spectrum can be represented by a theoretical model including atmospheric water vapor or by a flat spectrum model. We derive a 3-$\sigma$ upper limit of 1570~ppm on the presence of water vapor absorption in the atmosphere of XO-2b.} In an appendix, we perform a similar analysis for the gas giant planet XO-1b.
\end{abstract}





\section{Introduction}

Spectroscopy during planetary transits is a powerful tool to probe the atmospheres of exoplanets. The chemical composition and temperature profile can be derived, providing clues on the dynamics and chemical processes in the atmosphere. Molecular absorption in the atmosphere causes the planetary radius to vary with wavelength: the planet appears larger at wavelengths at which its atmosphere is more opaque. Thus, measuring the transit depth as a function of wavelength yields the spectrum of the planet's atmosphere. Typically, atmospheric components lead to variations of the transit depth of a few 100~ppm. The transit, when the planet passes in front of the star, probes the terminator region of the atmosphere. The eclipse, when the planet passes behind the star, yields the dayside spectrum of the atmosphere.

XO-2b is a 2.6 day period transiting hot Jupiter with a radius $0.996 \;\rm R_{jup}$ and mass  $0.565 \rm \;M_{jup}$ \citep{Burke2007}. The host star is an early K0 V dwarf of magnitude H = 9.34, radius and mass close to solar, and metallicity Fe/H = 0.45. A companion star is located at 31" separation (\simi 4600 AU at a distance d \simi 150 pc), and has an apparent magnitude H = 9.37, a metallicity Fe/H = 0.47, and a common proper motion $\mu$ = 157 mas yr$^{-1}$. Operationally, the presence of such a close-by twin companion provides an ideal comparison star.

Subsequent transit observations were conducted by \cite{Fernandez2009}, gathering 6 transits of XO-2b. By combining the lightcurve analysis with theoretical isochrones of stellar evolution models, parameters consistent with those of \cite{Burke2007} were found for the planet and host star. Only the period differs by 2.5~$\sigma$. This discrepancy can be explained by an additional body orbiting XO-2, or simply from an underestimation of uncertainties. 
Secondary eclipse measurements were taken on XO-2b with IRAC on the Spitzer Space Telescope \citep{Machalek2009}. They provide a tentative evidence for a weak temperature inversion in the upper atmosphere of the planet and a dayside temperature $T \sim 1500$ K, and predicted a substellar flux of $0.76\times10^9$~erg~cm$^{-2}$~s$^{-1}$.
Optical ground-based narrowband transit photometry with the 10.4~m Gran Telescopio Canarias yields clues on the atmospheric composition of XO-2b. Three transits were observed, each time in 4 wavelengths (679.2, 758.2, 766.5, and 883.9~nm), enabling detection of potassium in the planetary atmosphere \citep{Sing2011}. This atmospheric component, as well as sodium, was predicted to be a dominant source of opacity at optical wavelengths for close-in giant planets such as XO-2b \citep{Seager2000,Brown2001,Hubbard2001}. Models predict that the opacity from K, Na, and H$_2$ Rayleigh scattering dominate the optical, while H$_2$O absorption should dominate beyond 900~nm. Finally, the planet and host star parameters derived from GTC data agree well with \citet{Fernandez2009,Burke2007}, and best fits to atmospheric models are obtained using a planet temperature of 1500~K.

It has been proposed that hot Jupiters can be separated into two classes based on the incident flux received from their host star \citep{Fortney2008,Burrows2008}. The most irradiated planets, the pM class, are hot, harbor a temperature inversion due to the presence of TiO/VO gases in the upper atmosphere, and have a large day-night temperature contrast. The less irradiated planets, the pL class, are cooler, absorb flux deeper in the atmosphere, and the energy is redistributed better; they harbor no temperature inversion and have a lower day-night temperature contrast. The mixing ratio of TiO and VO in the atmosphere becomes smaller as the temperature decreases, leading atmospheric properties to change gradually between both classes. Given its substellar flux, XO-2b lies in this predicted transition region. Indeed, atmospheric potassium is a characteristic of pL class planets, whereas temperature inversion is expected for pM class planets. Models also predict that planets with a relatively low substellar flux would possess water features in absorption both on the dayside and at the terminator \citep{Fortney2008,Fortney2010}.

On a dynamical point of view, XO-2b belongs to the class of close-in giant planets spin-orbit aligned and orbiting prograde \citep{Narita2011}. Their radial velocity measurements are consistent with a null eccentricity. They identify a long term radial velocity variation in one of the epochs, which could be due to a hypothetical third body in the system.

The HST NICMOS observations of XO-2b presented here may be compared to those of four exoplanets, HD 189733b, HD 209458b, XO-1b, and GJ 436b. Water and methane were detected in the atmosphere of HD 189733b \citep{Swain2008}. Carbon monoxide was expected but not detected. The enhancement of methane instead of carbon monoxide thus brought a challenge to the understanding of chemical processes in the atmosphere of the planet, and possibly of other hot Jupiters. HD 189733b was again observed with NICMOS during an eclipse \citep{Swain2009b}. Water, carbon monoxide, and carbon dioxide were identified in the dayside spectrum of the planet, and the mixing ratio of these molecules was derived. 
Observations of the hot Jupiter HD 209458b revealed a dayside spectrum dominated by methane and water features, with smaller contribution from carbon dioxide \citep{Swain2009a}. Their analysis also confirmed the existence of a temperature inversion in the atmosphere of the planet. Several solutions were found plausible for the molecular abundances. 
NICMOS transmission spectroscopy was also used to probe the atmosphere of the hot Jupiter XO-1b \citep{Tinetti2010}. The extracted spectrum for XO-1b shows absorption features at a level of $\sim$300~ppm, interpreted as an atmosphere containing carbon dioxide, water, methane and possibly carbon monoxide.

However, several studies show that the amplitude of NICMOS systematics remaining after the analysis are comparable to the planet's expected signal. As an example, models for the hot-Neptune GJ-436b predict that water vapor will be present in the upper atmosphere at a level of 60~ppm to 110~ppm. From NICMOS observations, \cite{Pont2009} derive a flat spectrum with a RMS of 190~ppm. In particular, they find no significant absorption in the 1.4~$\mu$m water band; instead, they find a flux excess of 142~ppm. Given the noise level, the data do not constrain the models.
Another study of NICMOS transmission spectroscopy of HD 189733b, GJ-436b, and XO-1b finds no conclusive evidence for molecular features \citep{Gibson2011a}. Instead, residuals in the spectra are attributed to instrumental effects. The data themselves are the same as previous studies. The derived spectra are found to be dependent on the decorrelation procedure, necessary to remove instrumental noise. An answering comment from \citet{Deroo2010} emphasizes in particular that a good decorrelation can not be achieved without considering the angle of the spectral trace on the detector. Indeed when that parameter is omitted, as in the spectrum of XO-1b reported by \citep{Gibson2011a}, the result differs considerably from those in which it was included, \citep{Swain2008} and \citep{Tinetti2010}, as we illustrate in the Appendix. More generally, \citet{Gibson2011a,Gibson2011b} consider the linear decorrelation method used in previous studies inadequately removes systematics, and that the instrument model should not be extrapolated between orbits.

In this paper, we present near-infrared transit spectroscopy of XO-2b obtained with HST NICMOS. Particular attention is given to systematic effects in order to interpret features in the final spectrum. Fortunately, the XO-2 binary companion is close enough to appear simultaneously in the NICMOS field of view. The companion star does not have a transiting hot Jupiter, and is an ideal comparison star because its brightness and spectrum are almost identical to the star that hosts the transiting planet. This is the first time that a such a reference star is present in NICMOS exoplanet spectroscopic observations (a second star was present in the field of view of HD189733 \citep{Swain2008}, but was too faint to be useful). The analysis presented in this paper is systematically applied to both stars in the same way. This is a unique opportunity in NICMOS exoplanet spectroscopic studies to provide clues to the interpretation of results as molecular signatures of the planet's atmosphere or instrumental effects. 

The observations are described in Section~\ref{sec:Observations}, and the data reduction in Section~\ref{sec:Data reduction}. We present the white-light photometric analysis in Section~\ref{sec:White-light photometry}. The spectral analysis is detailed in Section~\ref{sec:Spectrum}, and the results are discussed in Section~\ref{sec:Discussion}. 

To better link this study to previous analysis of NICMOS planetary spectra, we re-analyze the XO-1b data following the procedure developed for XO-2b. The results are presented in appendix~\ref{sec:Analysis of XO-1 with NICMOS}.

\section{Observations}
\label{sec:Observations}

\begin{figure}[htbp]
   \centering
   \includegraphics[width=2.8cm]{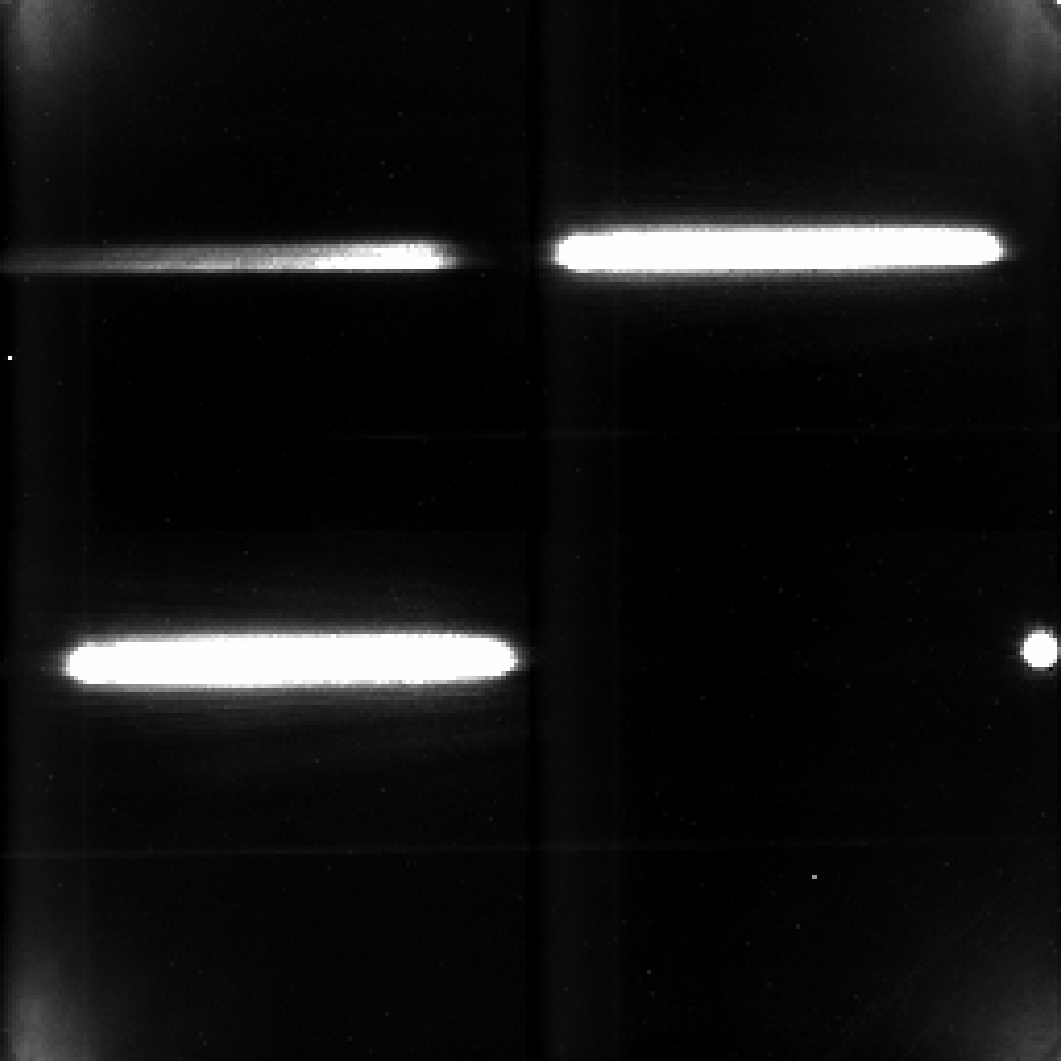}
   \includegraphics[width=2.8cm]{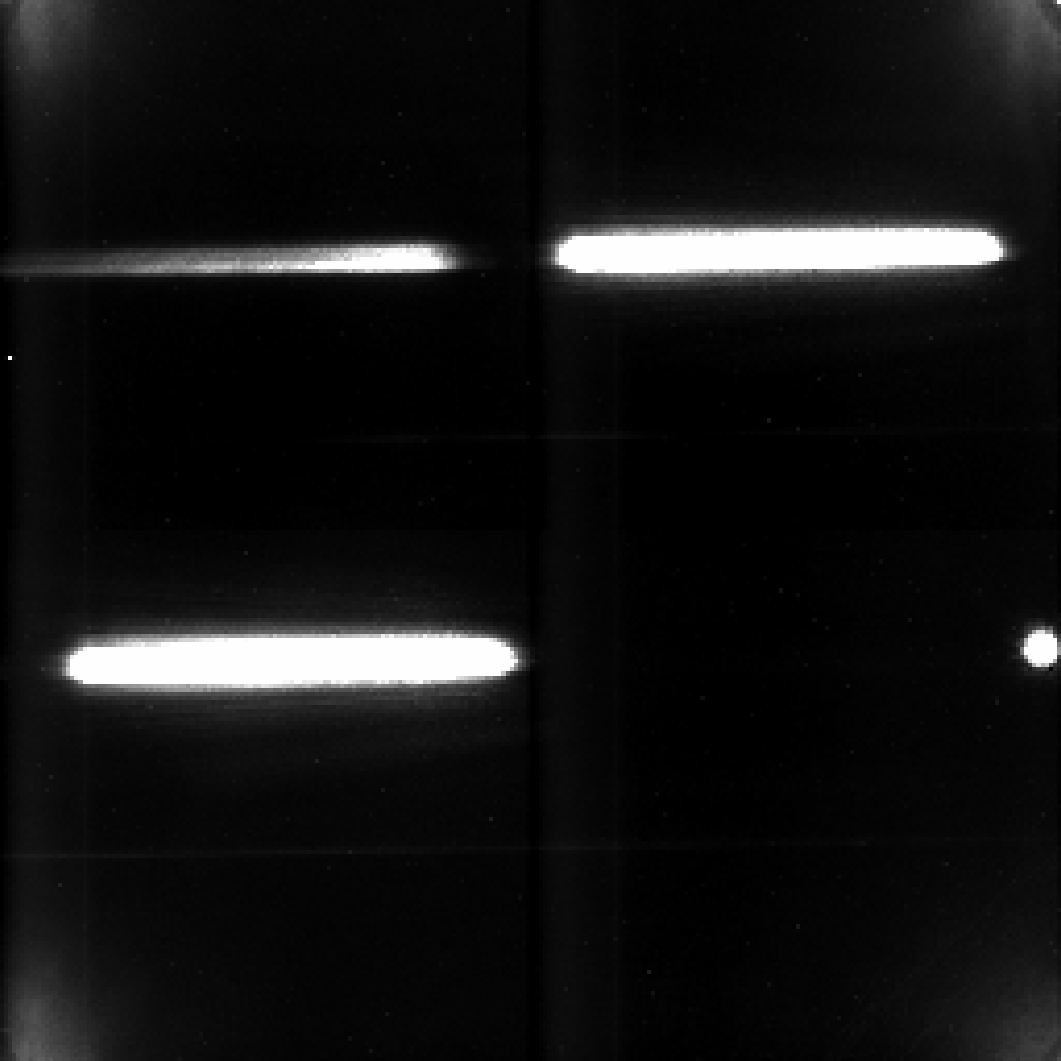}
   \includegraphics[width=2.8cm]{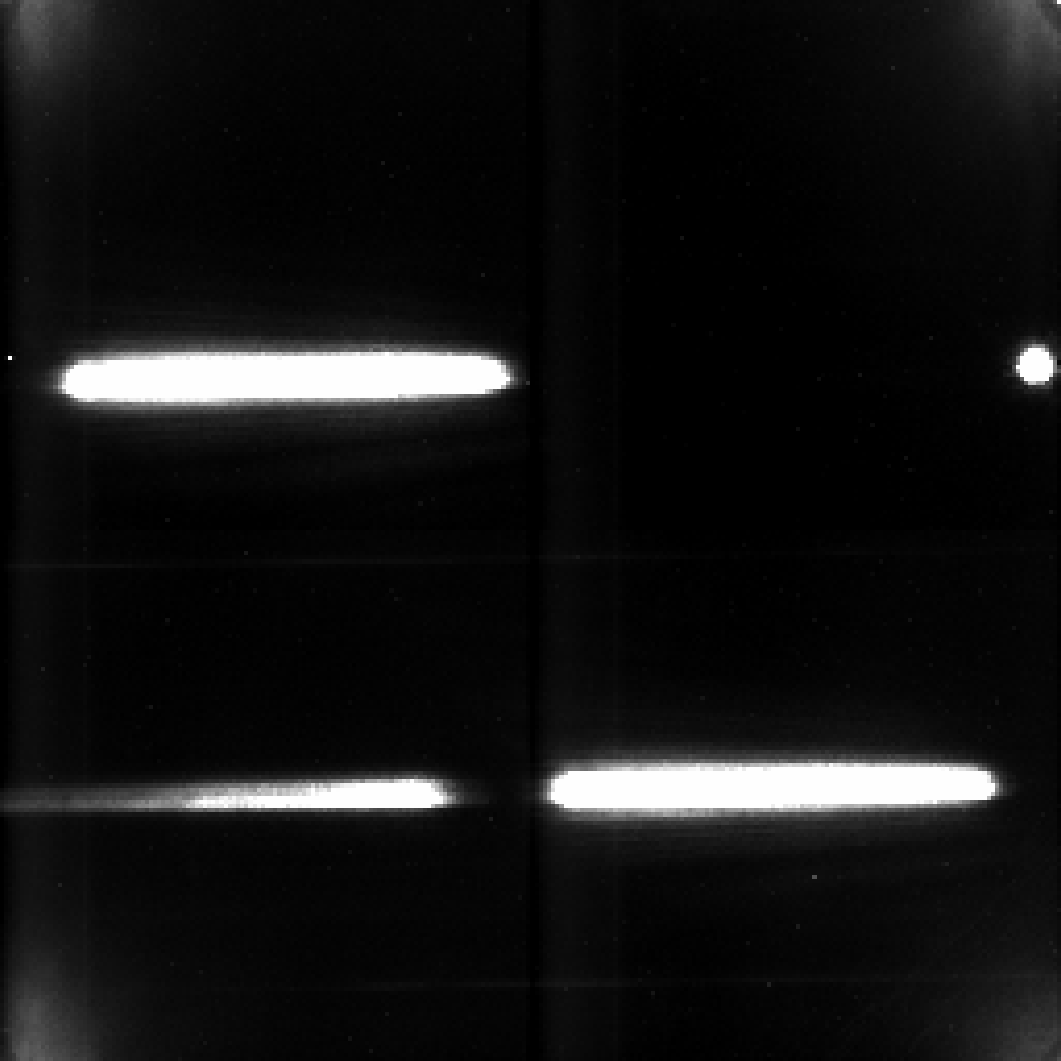}
      \caption{Position of the stars on the detector during the 1\st (left), 2\nd (middle), and 3\rd visit (right). XO-2~N is the upper star in the images. The images contain the 1\st order spectra of both stars, and the 0\tth order of XO-2~S for the 1\st and 2\nd visits and of XO-2~N for the 3\rd visit, due to a different orientation of HST.}
   \label{fig:images xo2}
\end{figure}

Three transits of XO-2b were observed in spectrophotometry by HST NICMOS, on 2007 November 2, 2007 December 11, and 2008 February 12. Each visit is divided into 5 HST orbits, containing 68 images of 32~s each. In total, 1006 images were acquired. We used the G141 grism, covering a spectral range from 1.1 to 1.9~$\mu$m. The length of the spectral trace is $\sim110$ pixels, and the PSF (Point Spread Function) was defocused to a FWHM (Full Width Half Maximum) of 5 pixels to minimize the effect of inter- and intra-pixel variations. The resulting spectral resolution $R=\lambda /{\Delta\lambda} = 37 $. The detector is read out in MULTIACCUM mode with NSAMP~=~10, \textit{i.e.} each image is composed of 10 successive readouts. When necessary, we will designate XO-2 as XO-2~N, and the companion star as XO-2~S (XO-2~N is located 30" north of XO-2~S). The position of both stars on the detector is the same in the 1\st and 2\nd visits, and different in the 3\rd visit (Figure~\ref{fig:images xo2}). The images always show the 1\st order spectrum of both stars, as well as the 0\tth order of XO-2~S for the 1\st and 2\nd visits and of XO-2~N for the 3\rd visit. A fainter 2\nd order spectrum is also visible.

\section{Data reduction}
\label{sec:Data reduction}

\subsection{Extraction of instrumental parameters}
\label{sec:Extraction of instrumental parameters}

\begin{figure*}[htbp]
   \centering
   \includegraphics[width=16cm]{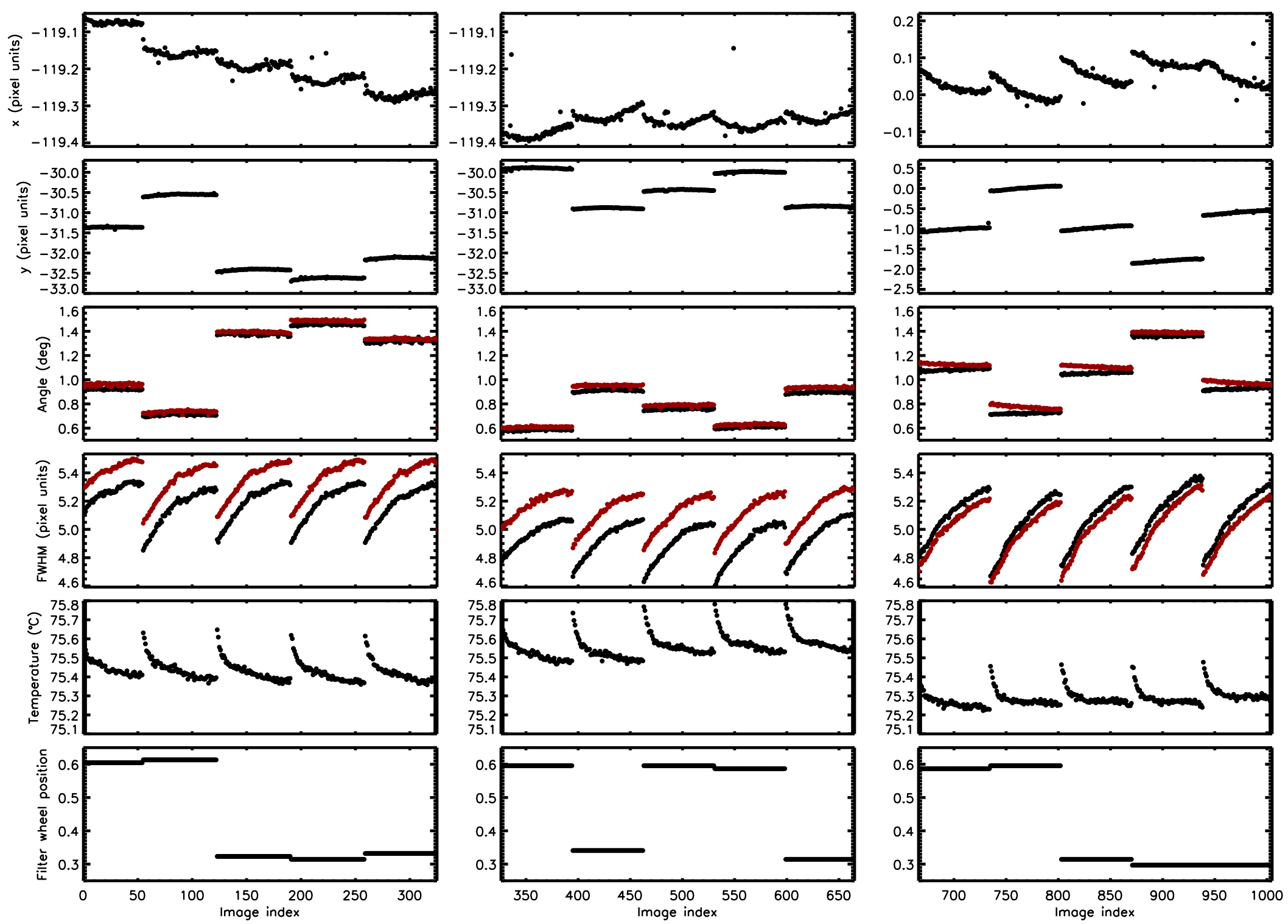}
      \caption{State parameters for the NICMOS XO-2 observations during the 1\st (left), 2\nd (middle), and 3\rd (right) visit. When relevant, the second star XO-2~S is indicated in red. From top to bottom: position in $x$ and $y$ of the 0\tth order present in the image, angle and FWHM of the spectral trace, temperature of the detector, and filter wheel position.}
   \label{fig:state parameters}
\end{figure*}

The instrumental parameters are derived for two reasons: to apply a wavelength-dependent flat-field correction, and to compute a decorrelation from these parameters in the derivation of the white-light transit curve and the planet spectrum. These parameters are represented in Figure~\ref{fig:state parameters}.

\begin{figure}[htbp]
   \centering
   \includegraphics[width=8cm]{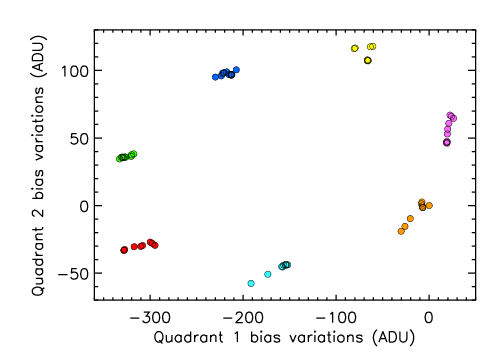}
      \caption{Variations of the bias level of quadrant 1 (lower left quadrant) versus quadrant 2 (lower right quadrant) for the 3\rd orbit of the 1\st visit, relatively to the bias level of the first image of the orbit. The 7 detector states appear clearly in the XO-2 data, and are represented by different colors.}
   \label{fig:get7phases}
\end{figure}

\begin{itemize}

\item Position $(x, y)$ of the 0\tth order.\\
We derive the $(x, y)$ position of the observed 0\tth order by cross-correlation with a reference image. The 0\tth order $(x, y)$ position of the other star, which does not appear in the image, is derived by applying an offset to the 0\tth order that is present. This offset is calculated by correlating the 1\st order spectral traces of both stars, and considering that the same shift applies to the 0\tth orders. The best correlation is obtained by applying a constant shift of 118 pixels in the $x$ direction and 99 pixels in the $y$ direction for visit 1 and 2 for XO-2~N with respect to XO-2~S, and a shift of 118 pixels in the $x$ direction and $-99$ pixels in the $y$ direction for visit 3 for XO-2~S with respect to XO-2~N.

\item Angle and FWHM of the spectral trace.\\
The angle and FWHM of the 1\st order spectral trace are derived for each star. The spectral trace is divided into boxes of 5 pixels wide in $x$ and 40 pixels wide in $y$. Each box is averaged in $x$ by a median filter, and a 1-dimension Gaussian fit is computed in the $y$ direction. A linear fit of the center of all Gaussians yields the angle of the spectral trace. Each Gaussian also provides a FWHM in $y$, and the median of all values gives the FWHM of the spectral trace.

\item Detector temperature and state.\\
The detector temperature is derived from the bias level using a standard procedure for NICMOS \citep{Pirzkal2009b}. In addition, \citet{Burke2010} showed that the detector electronics vary between 7 states, resulting in photometric variations approximately at the millimagnitude amplitude, peak to valley (Figure~\ref{fig:get7phases}). The detector state in a given image is inferred by comparing the bias level in two different quadrants. These 7 states are clearly identified in the XO-2 data.

\item Filter wheel position.\\
The filter wheel has a constant position during an orbit, but experiences small shifts between orbits due to the blank filter element insertion during Earth occultation of the target. These variations are believed to affect the position of the spectral trace, in particular its angle \citep{Gilliland2003}. The filter wheel as 2 positions.

\item HST phase.\\
The HST orbital phase is known to influence NICMOS data. However, this parameter is responsible for variations of other parameters that are more directly linked to the images, such as the FWHM and the detector temperature \citep{Wiklind2009}. 

\end{itemize}

\subsection{Bias calibration}

The calibration is made from the raw images. The bias calibration is performed by subtracting the first readout from the last \citep{Fixsen2000}. Pixels with a dark current greater than 1\elec/s are already marked as bad, and lower dark currents are negligible for our purpose.

\subsection{Bad pixel correction}

To correct for bad pixels, we combine the nominal bad pixel mask for the NIC3 detector with the pixel flags provided in each individual image. In addition, we flag as bad pixels those that deviate by more than 3.5 $\sigma$ from their average value in the orbit, or from their value in the adjacent images. We find typically 35 bad pixels out of $\sim2100$ in the spectral aperture. Each bad pixel's value is then replaced by a spline interpolation of adjacent pixels.

\begin{figure}[htbp]
   \centering
   \includegraphics[width=4cm]{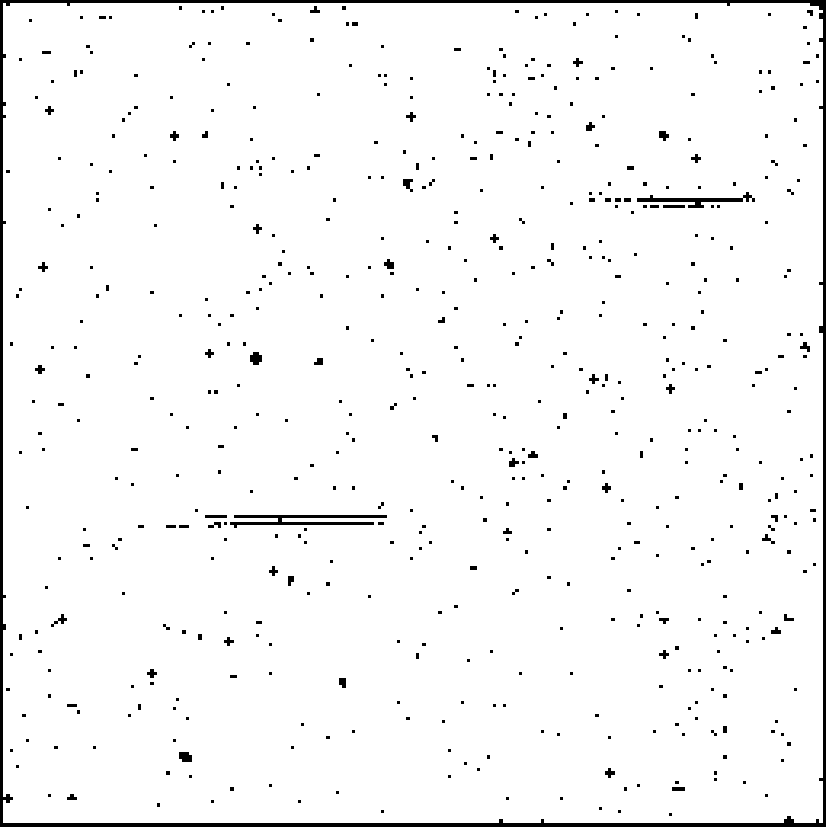}
   \includegraphics[width=4cm]{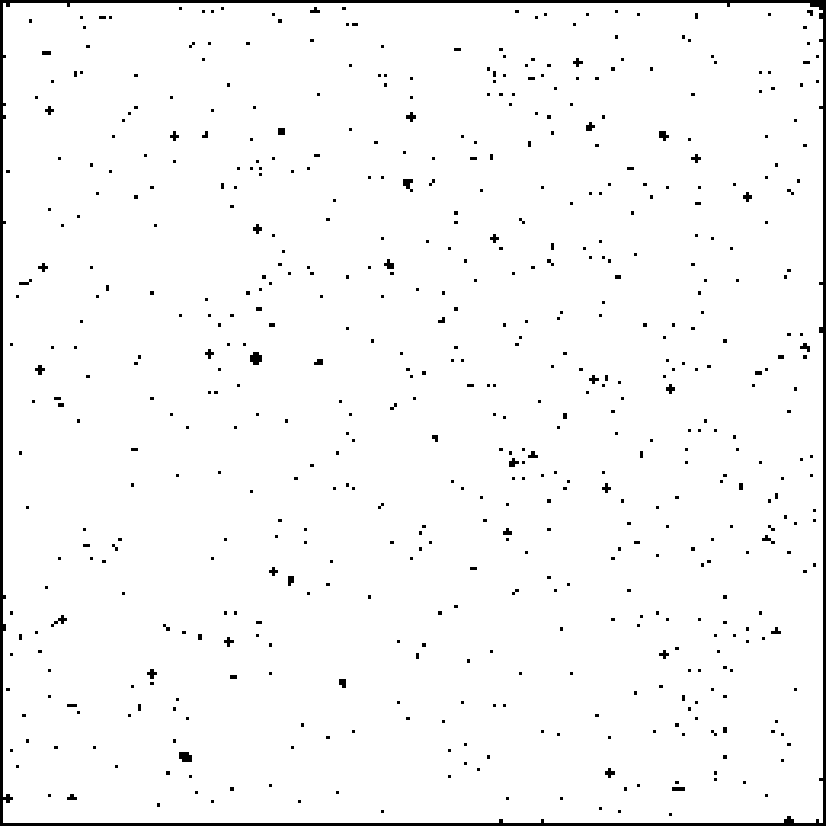}
      \caption{Example of bad pixel masks at the beginning of an orbit (left), and after the instrument is stabilized (right). Lines at the location of the spectral traces appear at the beginning of orbits.}
   \label{fig:bad pixels mask}
\end{figure}

The set of bad pixels for each image defines a mask (Figure~\ref{fig:bad pixels mask}). In some masks, lines appear along the $x$ axis at the location of the spectral traces. This behavior appears mostly at the beginning of the orbits, for which the detector temperature is significantly higher and the FWHM significantly smaller than during the rest of the orbit, due to the satellite stabilizing after re-entering the Earth shadow (Figure~\ref{fig:state parameters}). These lines are thus due to a different flux distribution compared to the rest of the orbit, and not to bad pixels. Typically, the first 20 images of each orbit are affected, plus \simi12 other images in each orbit of the 3\rd visit. In total, this represents 37\% of the data. Since this fraction is large, we process these images similarly to the rest of the data, without considering those lines as bad pixels. The decorrelation from instrumental parameters will reduce the effects of such variations.

\subsection{Flat-field correction}

Since the response of the pixels is wavelength dependent, a custom-made flat-field is required.\footnote{Note that wavelength-dependent flat-fielding cancels out at first order for exoplanet spectroscopy, because the transit depth is obtained by comparing in-transit to out-of-transit orbits at each particular wavelength. The out-of-transit data, in effect, flat fields the in-transit data, except potentially for non-linearities in the detector or analysis technique.}
For each image, we calculate the reference position of the star using an image taken without the dispersive element. This reference position and the beam angle are used to calculate the wavelength at each pixel, using the grism's nominal dispersion relation \citep{Pirzkal2009a}. Then, the response of each pixel as a function of wavelength is determined by evaluating a 4\tth-order polynomial to flat-fields taken at 10 different wavelengths. A flat-field is made for each image, and an average flat-field is built for each visit. The latter one is used for the calibration. Images of a given visit are thus calibrated using the same flat-field. 

\section{White-light photometry}
\label{sec:White-light photometry}

\subsection{Lightcurve extraction}
\label{sec:Lightcurve extraction}

In order to derive the star and planet parameters, we perform white-light photometry by summing the flux over the spectral trace. We optimize the photometric apertures for the star and sky background. We extract and compare lightcurves for apertures of different sizes, varying between 6 and 50 pixels in $y$. The sky background is computed from the same quadrant as the spectral trace, either above or below depending on the position of the star; the width of the sky region varies between 40 and 128 pixels in $x$, and between 20 and 60 pixels in $y$. The sky level is calculated with an outlier resistant mean with a threshold at 3.5 $\sigma$, multiplied by the number of pixels in the star region, and subtracted to the stellar flux. The lightcurve RMS is calculated from the 2\nd and the 5\tth orbit of each visit, to avoid the first orbit which is commonly affected strongly by systematics, as well as the ingress, egress, and in-transit orbits. Independently for each visit, we determine the optimal height in $y$ as the one resulting in the smallest RMS. We find an optimal height in $y$ of 18, 20, and 18 pixels for XO-2~N, and 24, 18, and 16 pixels for XO-2~S, for the 1\ste, 2\nde, and 3\rd visit respectively.

\begin{figure}[htbp]
   \centering
   \includegraphics[width=8cm]{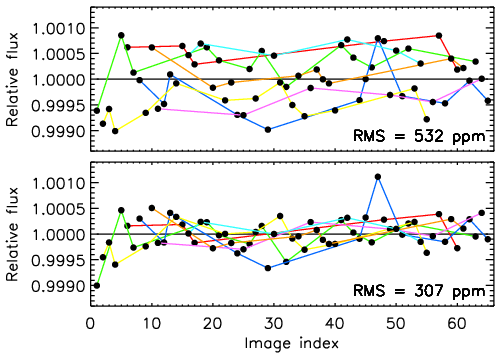}
      \caption{Relative flux of XO-2~N as a function of the image index for the 2\nd orbit of the 2\nd visit, before (top) and after (bottom) correction of the systematic offset induced by the 7 detector state (astrophysical variation due to the transit has been corrected prior to display). Images sharing the same detector state are connected by lines using the same color code as in Figure~\ref{fig:get7phases}. Measurements at each state are systematically offset from the mean flux during the orbit. Here, the correction lowers the RMS from 532 to 307~ppm.}
   \label{fig:cor7phases}
\end{figure}

The RMS is calculated after correcting the orbit lightcurves for the 7 detector states (Figure~\ref{fig:get7phases}), without applying any other decorrelation. Indeed, the detector state clearly affects the photometry: flux measurements corresponding to a given state are systematically offset from the mean flux during the orbit. Following the procedure outlined in \cite{Burke2010}, a correction is applied by correcting the mean of the measurements taken in each state to the mean flux of the orbit, after accounting for the transit if present. The 7-state correction lowers the RMS by a factor 1.5 on average (Figure~\ref{fig:cor7phases}). Because orbits have a different average flux, the correction cannot be applied for a whole visit at once when extracting these raw lightcurves. However, in the remainder of this work, the average flux of each orbit is adjusted by optimization processes, and the 7-state correction will be applied for each visit instead of each orbit. Using the larger number of data points in a visit, compared to each orbit, provides better estimates of the seven state-dependent flux offsets and also reduces to insignificance the side-effect that fitting these offsets will reduce even random noise simply because each offset is another parameter in the fit.

\subsection{Monte-Carlo analysis}
\label{sec:Monte Carlo analysis}

The star and planet parameters are derived through a Markov Chain Monte-Carlo (MCMC) minimization process \citep{Burke2007}. We simultaneously fit the lightcurve with a transit model and an instrumental model. We adopt a quadratic limb-darkened transit model from \citet{Mandel2002}. The model parameters, some of which are degenerate, are the period $P$, the stellar mass $M_\star$, the stellar radius $R_\star$, the planet radius $R_p$, the inclination $i$, the mid-transit time for each visit $t_1$, $t_2$, $t_3$, and the limb-darkening coefficients $u_1$, $u_2$. For an efficient MCMC calculation, we remove degeneracy between parameters by using $R_\star$, $r=R_p/R_\star$, and the transit duration from first to fourth contact $\tau_{14}$, instead of $R_\star$, $R_p$, and $i$. For similar reasons, we use $a_1=u_1+2u_2$ and $a_2=2u_1-u_2$, instead of $u_1$ and $u_2$ (Holman et al. 2006). We fix $P$ and $M_\star$, the other parameters are free. We assume zero eccentricity, as indicated by radial velocity measurements \citep{Burke2007}. Initial values are taken from \citet{Sing2011} and \citet{Fernandez2009}, except for the limb-darkening coefficients. These are derived in the I, J, H, and K bands from tabulated values \citep{Claret2000} using the stellar mass, temperature and metallicity found by \citet{Burke2007}, and interpolated at the mean of the NICMOS band, 1.54~$\mu$m. We find $u_1=0.054$ and $u_2=0.417$.

The minimization process includes a decorrelation from instrumental parameters. The instrumental model includes a second order polynomial in the angle $\alpha$, and a linear combination of $x$, $y$, the FWHM $w$, and the detector temperature $T$. The 3 visits are treated separately. In addition, the filter wheel is displaced after HST re-enters the Earth shadow leading to 2 distinct positions (see Figure~\ref{fig:state parameters}). Thus, we divide the data into 6 groups, with 2 groups per visit. A multiplicative constant $F_0$ is used to adjust the flux level between each group. The instrumental model $\Psi$ is computed independently for each group $g$, whereas the transit model $\Phi$ remains the same. The model lightcurve thus can be written as:

\begin{equation}
F_{mod} \; = \;  \Phi(R_\star,r,\tau_{14},t_1,t_2,t_3,a_1,a_2)  \;\times\big\{F_{0}\times\Psi(\alpha^2,\alpha,x,y,w,T)\big\}_{g \,\in\,\llbracket 1,6 \rrbracket}
\label{eq:mcmc photometry}
\end{equation}

The first orbit of each visit is strongly affected by HST settling; those data are discarded. The time for each observation is the midpoint of the exposure converted to barycentric Julian date. The priors for the parameters are similar to those defined in \cite{Burke2007, Burke2010}: equivalent to uniform on $R_\star$, $R_p$ , and $i$ for $R_\star$, $r$, and $\tau$, and uniform for the other parameters. A correction for the 7 readout states is applied during the process, using the known detector state at each image as calculated section \ref{sec:Extraction of instrumental parameters}. The flux offset generated by each state is calculated using the whole visit. The likelihood is modeled as independent Gaussian residuals with uncertainty adjusted from two trial runs to enforce a final reduced $\chi^2$ of 1. The value $\sigma= 380$~ppm is adopted, and remains fixed during the MCMC run. This uncertainty is treated as white noise. We adopt the median as the best estimate of the posterior probability, and the 1$\sigma$ uncertainties of the parameter estimates are determined such that they demarcate 68.3\% of the MCMC samples. The first 1000 iterations are discarded as burn-in period of the MCMC procedure. 

Our choice of the parameters for the instrumental model is guided by the data. In many trial parameter sets, we examine sets including $F_0$, $x$, $y$, $\alpha$, $w$, $T$, the HST orbital phase $\phi$, and quadratic and cubic powers of $x$, $y$, and $\alpha$. We perform short runs of $10^4$ iterations for each set of parameter combinations. In particular, we evaluate the importance of each parameter in a simple iterative process: one parameter is added each time, and the one that yields the lower RMS is retained, if the RMS is lowered by at least 10 ppm. \modif{(this arbitrary choice constitutes a limitation of the method)}. Examples of combinations are given in Table~\ref{tab:parameter combinations}. This empirical approach is useful to understand the NICMOS systematics. We find that $F_0$ is the most important parameter. The average flux varies by up to 0.6\% between groups. These variations are also present in previous studies of NICMOS data \citep{Burke2010,Gibson2011a}, and their origin is not understood. Incidentally, $F_0$ also accounts for stellar variability between visits. The second most important parameter is the angle of the spectral trace, $\alpha$. In decreasing order of effect, we then find $x$ and $y$, then $w$ and $T$. Among higher order terms, only $\alpha^2$ yields a significant improvement. Including the HST orbital phase yields a negligible improvement and is not included. As seen in Figure~\ref{fig:state parameters}, the HST orbital phase is in fact responsible for variations of the actual instrumental parameters $w$ and $T$, which are already accounted for. After examining a large number of possible combinations, we adopt the parameters expressed in Equation~\ref{eq:mcmc photometry}.

\begin{figure}[htbp]
   \centering
   \includegraphics[width=11cm]{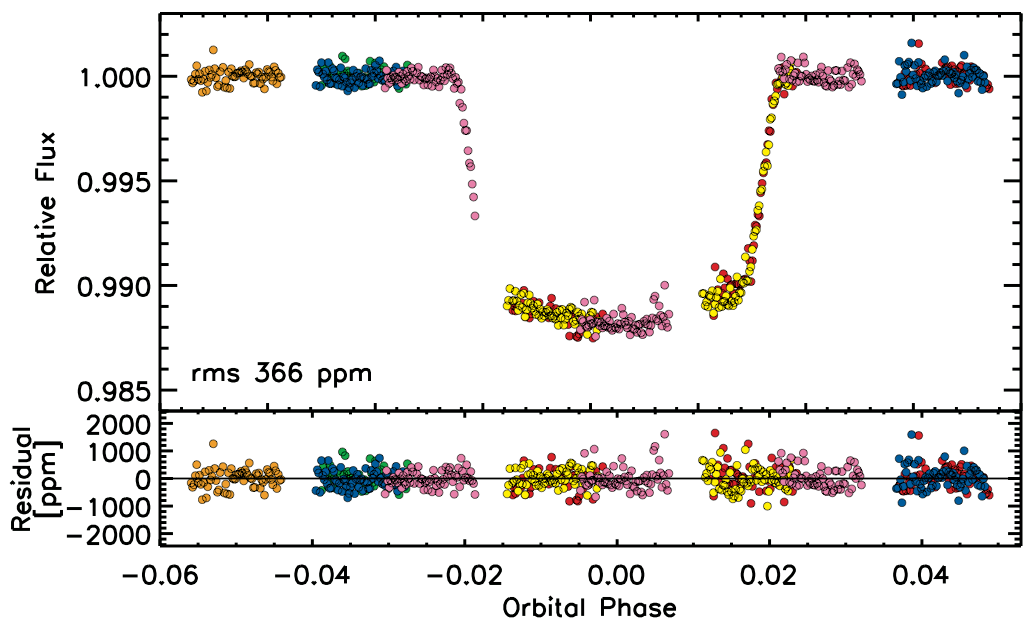}
      \caption{White-light curve (top) and residuals (bottom) for \mbox{XO-2~N} after the MCMC analysis. Different colors represent the 6 groups of data. The best fit model yields a point-to-point precision of 350~ppm.}
   \label{fig:white-light curve}
\end{figure}

To derive the star and planet parameters, we run 5 MCMC chains of $10^5$ iterations and merge their posterior probability distributions. The results are reported in Table~\ref{tab:final param mcmc}. The best fit model yields a point-to-point precision of 366~ppm (Figure~\ref{fig:white-light curve}). The Poisson noise limit is 203~ppm per exposure. The minimum $\chi^2$ is 716 for 719 degrees of freedom, indicating uncertainties are well estimated.

\begin{table}[htdp]
\begin{center}
\caption{Examples of white-light curve precision after decorrelation from different instrumental parameter combinations.}
\begin{tabular}{lc}
\\
\hline
\hline
  Instrumental parameters &  Precision (ppm)   \\
\hline

None & 2790   \\
$F_0$ & 625   \\
$F_0$, $\alpha$ & 560   \\
$F_0$, $\alpha$, $x$, $y$ & 423   \\
$F_0$, $\alpha$, $x$, $y$, $w$, $T$ & 409   \\
$F_0$, $\alpha$, $x$, $y$, $w$, $T$, $\alpha^2$ & 367   \\

\hline  
\label{tab:parameter combinations}
\end{tabular}
\end{center}
\end{table}

\begin{table}[htdp]
\begin{center}
\caption{Parameters for the XO-2~system.}
\begin{tabular}{ccccc}
\\
\hline
\hline
   Parameter &  Units & Value & \multicolumn{2}{c}{1-$\sigma$ uncertainty}   \\
\hline

$P$     & [days]   &       2.61586178$^\dagger$ &  &  \\
$M_{\star}$    & [$\rm M_{\odot}$]   &       0.971$^\ddag$   &  &  \\           
$R_{\star}$     & [$\rm R_{\odot}$]   &       0.990    &    $-$  0.009 &  $+$  0.009       \\
$R_p/R_{\star}$    &    &       0.10304    &    $-$ 0.00037    &   $+$  0.00037 \\
$t_1$     & [HJD]   &     2454406.71987      & $-$   0.00016     &  $+$   0.00017  \\
$t_2$     & [HJD]   &     2454445.95817      & $-$   0.00016     &  $+$   0.00019 \\
$t_3$     & [HJD]   &     2454508.73829      & $-$   0.00016     &  $+$   0.00014 \\            
$\tau_{14}$  & [hours]   &       2.6839    &   $-$ 0.0077     &  $+$ 0.0066   \\
$b$   &    &       0.28    &   $-$  0.044    &   $+$  0.036 \\  
$i$    & [deg]   &      88.01    &   $-$  0.28    &  $+$   0.33 \\
$a$   & [AU]     &	      0.0368 	&  &   	  \\ 
$a/R_{\star}$    &    &       7.986    &  $-$   0.073    &   $+$  0.074 \\
$\rho_{\star}$   & [g cm$^{-3}$]   &       1.410    &  $-$   0.039    &  $+$   0.040 \\ 
$log \; g_{\star}$   & [cm s$^{-2}$]   &       4.433    &  $-$   0.008    &   $+$  0.008 \\
$u_1$     &    &       0.359    &   $-$  0.045    &  $+$   0.042 \\
$u_2$     &    &      $-$0.029   &   $-$  0.071    &  $+$   0.077 \\
$M_p$     & [$\rm M_{jup}$]   &       0.566$^*$    &   $-$  0.025    &   $+$  0.025 \\
$R_p$     & [$\rm R_{jup}$]   &       0.993    &   $-$  0.012    &  $+$   0.012 \\
$\rho_p$  & [g cm$^{-3}$]   &       0.715    &   $-$  0.040    &   $+$  0.041 \\ 
$log \; g_p$   & [cm s$^{-2}$]   &       3.152    &  $-$   0.022    &  $+$   0.022 \\
$\Theta$    &     &       0.043    &   $-$  0.002    &   $+$  0.002 \\

\hline  

\label{tab:final param mcmc}
\end{tabular}
\end{center}

\vspace{-0.5cm}
$^\dagger$ From \citet{Sing2011} \\
$^\ddag$ From \citet{Fernandez2009} \\
$^*$ Using radial velocity measurements from \citet{Burke2007} \\
$\Theta$: Safronov number

\end{table}

\section{Spectrum}
\label{sec:Spectrum}

\subsection{Instrumental parameters that affect the spectrum}
\label{sec:Identifying NICMOS noise sources}
 
\begin{figure}[htbp]
   \centering
   \includegraphics[width=8cm]{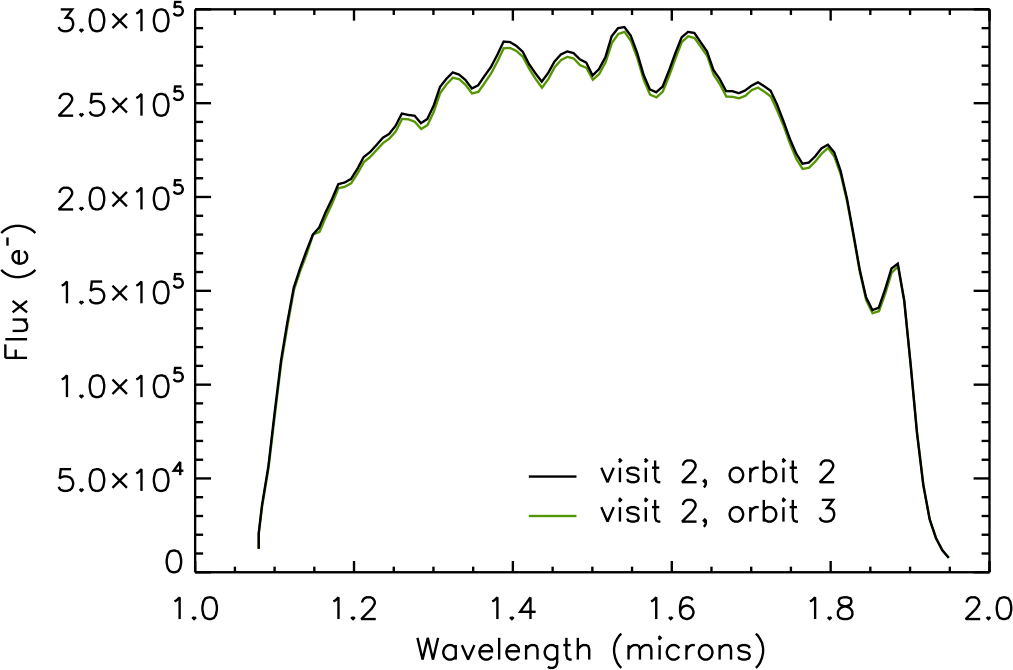}
      \caption{Example of raw spectrum obtained for the out-of-transit 2\nd orbit (black) and the in-transit 3\rd orbit (green) of the 2\nd visit. The latter is naturally shifted in flux by an amount corresponding to the transit depth. Variations with wavelength are due to the transmission response of the grism.}
   \label{fig:rawspec}
\end{figure}

To identify the main sources of systematics affecting the spectrum, we first build the XO-2~Stellar spectrum in a simple way: on each image, the flux is summed over each column of the spectral trace, after subtracting the sky background in that column. The bounds in the $y$ direction of the spectral trace and the sky region are the same as in the white-light analysis Section~\ref{sec:Lightcurve extraction}. These spectra are affected by strong variations due to the grism response in wavelength: for example, 15\% between 1.54 and 1.58~$\mu$m (Figure~\ref{fig:rawspec}).
We estimate the influence of the sky background. The median sky background is 27 \elec/px, which is $5.4\times10^{-4}$ lower than the median flux pixels inside the stellar region. A determination of sky background in the whole image rather than in each column leads to no significant changes. 

\begin{figure}[htbp]
   \centering
   \includegraphics[width=7.7cm]{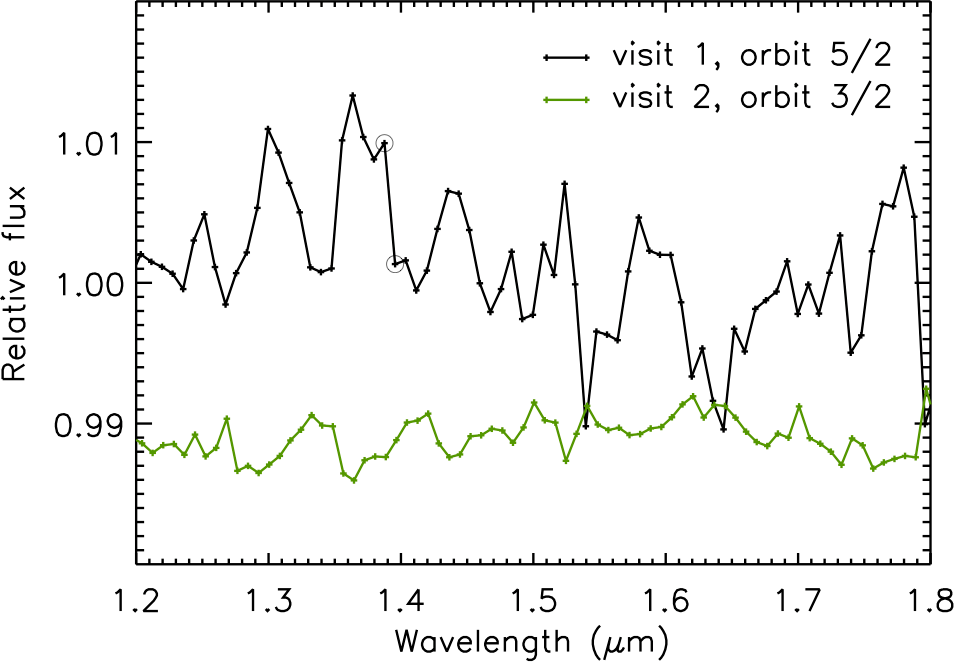}
      \caption{Example of ratio of spectra obtained by dividing the out-transit 5\tth orbit by the out-of-transit 2\nd orbit of the 1\st visit (black), and the in-transit 3\rd orbit by the out-of-transit 2\nd orbit of the 2\nd visit (green). The latter is naturally shifted in flux by an amount corresponding to the transit depth. Variations in those spectra are respectively 2.3 and 0.6\% peak-to-peak. Circled points are discussed in the text.}
   \label{fig:rawspec-div}
\end{figure}

A common method to extract the planetary spectrum is to divide an in-transit by an out-of-transit stellar spectrum. To empirically infer the uncertainties of this method for our data, we use all possible pairs of orbits for each star and each visit including pairs of out-of-transit orbits, but as before, discarding the first orbit of each visit. We do not use the ingress and egress. The resulting spectra have strong variations, comprised between 0.3 and 3\% peak-to-peak (Figure~\ref{fig:rawspec-div}). For comparison, the precision of the Poisson noise limited spectrum would be 370~ppm. Evidently, the large variations are due to systematic effects.

\begin{figure}[htbp]
   \centering
   \includegraphics[width=8cm]{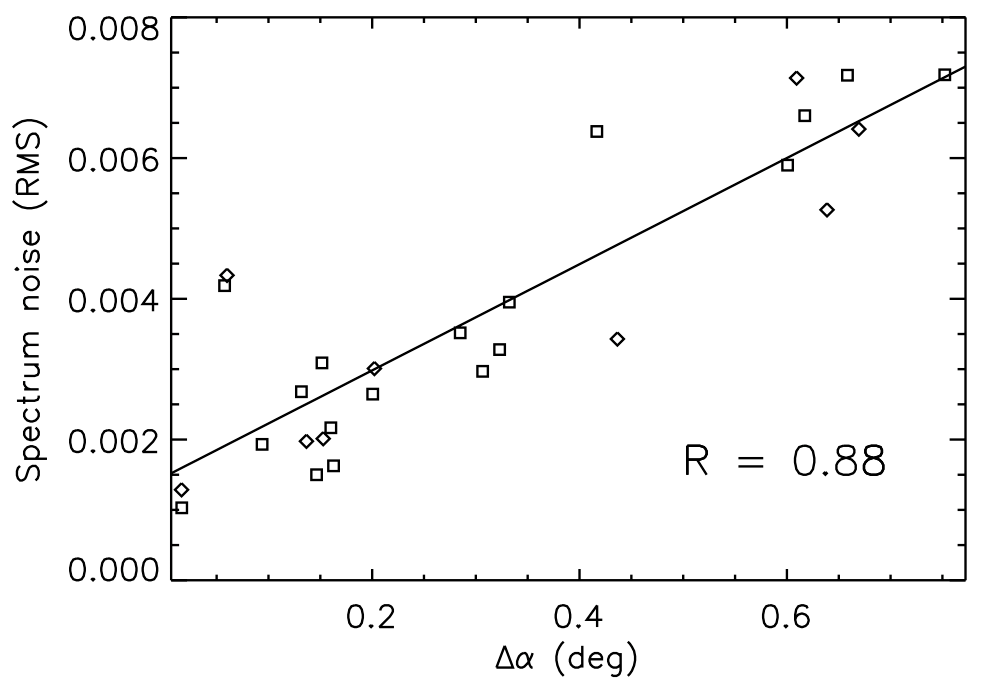}
      \caption{Correlation of the RMS of the ration of spectra with the difference $\Delta\alpha$ of the angle between the two orbits. Each point corresponds to a given pair of orbits. Results for XO-2~N are represented with squares, and results for XO-2~S with diamonds; they follow the same trend. The linear fit has a correlation coefficient R=0.88.}
   \label{fig:correlation rawspec alpha}
\end{figure}

Similar to our analysis of the white-light curve, we investigated correlations of the spectrum with instrumental parameters. We calculate the variation of a given parameter between two orbits by comparing its mean value in these orbits. We find a clear correlation with $\alpha$: a larger difference in $\alpha$ yields a larger RMS variation in the ratio of two spectra (Figure~\ref{fig:correlation rawspec alpha}). The 0.6\% at $\Delta\alpha = 0.6^\circ$ is consistent with 10\% intrapixel variations for in-focus G141 spectra (Bohlin R., private communication) diluted by the defocus of 5 px we intentionally introduced specifically to reduce the effect of intrapixel sensitivity variations \citep{Lauer1999}. Angle variations and more generally position variations combined to intrapixel variations apparently are a significant source of NICMOS systematics.

\begin{figure}[htbp]
   \centering
   \includegraphics[width=7cm]{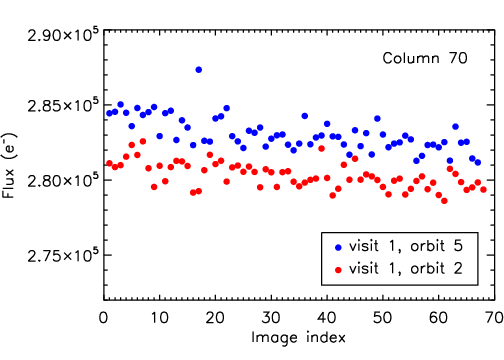}
   \includegraphics[width=7cm]{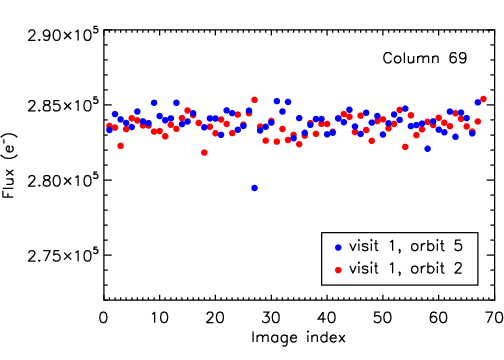}
      \caption{Lightcurves of columns 70 (left) and 69 (right) during the 5\tth (blue) and 2\nd (red) orbits of the 1\st visit, corresponding respectively to the left and right circled points in Figure~\ref{fig:rawspec-div}. A variation between these 2 orbits appears clearly for column 70, and not for column 69.}
   \label{fig:lightcurves-col-69-70}
\end{figure}

To investigate the variations in the spectra, we analyze in detail the light curves of each column. We find that the noise in the light curves is $\sim1.5$ times the Poisson noise within a single orbit. In contrast, global variations appear between different orbits. As an example, Figure~\ref{fig:lightcurves-col-69-70} illustrates the light curves of columns 70 and 69 during the 2\nd and 5\tth orbits of the 1\st visit. Those columns correspond respectively to the left and right circled points in Figure~\ref{fig:rawspec-div}. A systematic flux difference between the 2 orbits appears clearly for column 70, and not for 69. Because this variation differs for each column, it is not due to a white-light flux variations. Moreover, in our example, the white-light flux differs by 650~ppm between both orbits, whereas the mean of the column 70 lightcurve differs between orbits by \simi15 times this value. This variation is thus due to more complicated systematic effects. The same behavior appears when extracting the lightcurves directly from the raw images, and is therefore present in the original data.

Images do not show any particular feature that could result in a systematic flux difference. Only the level of each individual pixel differs slightly between orbits. Considering, for example, the 30\tth image of both orbits, this results in a flux variation of 900~ppm for column 69, and in a variation 10 times greater for column 70. These differences at the pixel level can typically be caused by intrapixel variations combined with motion of the spectral trace between orbits. We conclude that this is a main limitation of NICMOS for the extraction of planetary spectra.

\begin{figure}[htbp]
   \centering
   \includegraphics[width=12cm]{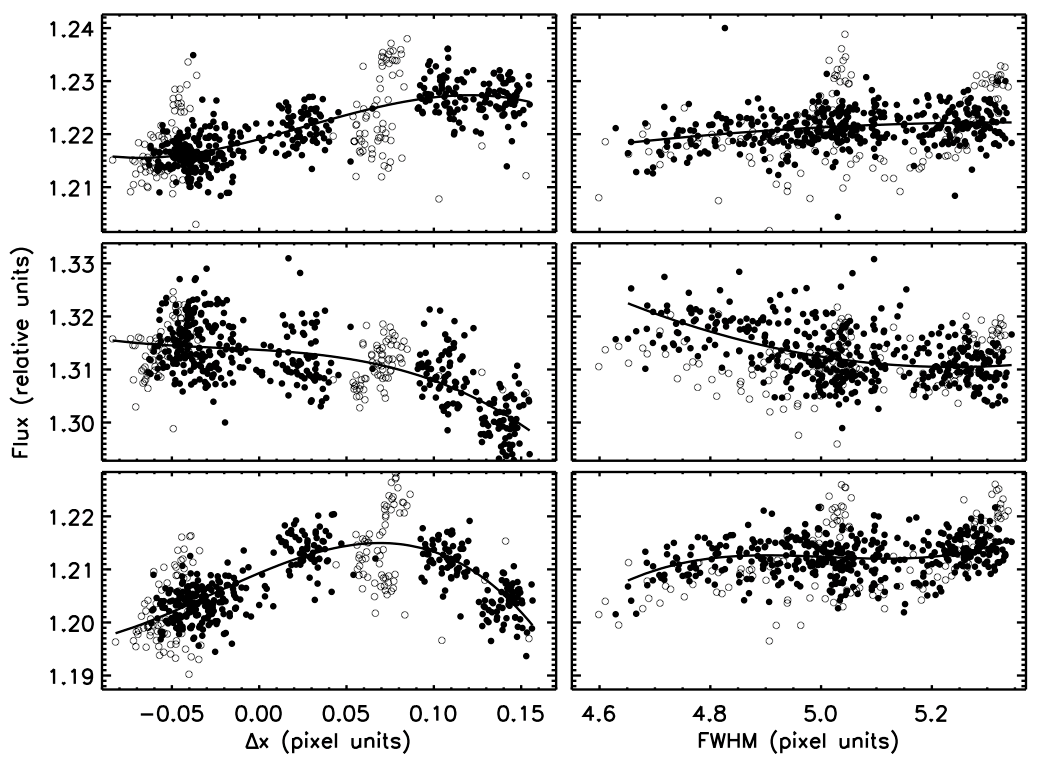}
      \caption{Correlation of the lightcurves of columns 49 (top), 71 (middle), and 75 (bottom) with the $x$ position (left), and the FWHM (right) for the 1\st and 2\nd visits together. A 3\rd order polynomial fit is shown (plain line). Lightcurves are first normalized to the mean white-light flux of the orbit. Open circles represent data taken during ingress and egress, not used for the fit.}
   \label{fig:correlations-1param}
\end{figure}

Correlations with other parameters are also identified in the column lightcurves, as shown Figure~\ref{fig:correlations-1param}. A correlation with the $x$ position appears for most of the columns, which may be caused by the variations of the grism response in wavelength. Indeed, the observed $\sim$1\% variations in the lightcurves is consistent with grism response variations up to 3\%/px and a measured $\Delta x=0.3$ px between the 1\st and 2\nd visit. Intrapixel variations may also play a role. We also find a frequent correlation with the FWHM. Similar correlations appear for XO-2~S. Note correlations such as shown Figure~\ref{fig:correlations-1param} may be the result of multi-dimensional correlations projected onto a single-parameter axis. Removing 1-dimensional correlations by polynomial fit reduces the noise in the spectra to $\sim0.1\%$ (1000~ppm). Another factor of \simi10 is necessary to extract the planetary signal. A more sophisticated decorrelation method is required, as described in the next section.

The apparent non-linear variations may be the result of a multi-parameter dependence projected on a single-parameter axis, and we keep the parameters defined in section \ref{sec:Monte Carlo analysis} for the decorrelation section \ref{sec:Spectrum Monte Carlo analysis}.

\subsection{Monte Carlo analysis}
\label{sec:Spectrum Monte Carlo analysis}

A Monte-Carlo analysis is performed to derive the spectrum of XO-2b. The columns are binned by 5 (the size of the FWHM), giving 22 independent channels along the spectral trace. The corresponding spectral resolution is R=37. To aid comparison with other NICMOS analysis \citep{Swain2009b,Tinetti2010}, each bin is then shifted by 0, 1, 2, 3 and 4 px, leading to as many points as the number of columns ; the analysis is performed independently for each of the $22\times5$ channels. An alternative approach to MCMC would be to use Gaussian processes to account for systematics, as done by \citet{Gibson2012a,Gibson2012b}.

A lightcurve is extracted for each channel using a box of 5 px wide in $x$ and bounds in $y$ that were found optimal in Section~\ref{sec:Lightcurve extraction}. We exclude the first orbit of each visit and individual points departing by more than 3 $\sigma$ from the mean flux of the orbit. These lightcurves are then fitted with a Monte-Carlo procedure similarly to Section~\ref{sec:Monte Carlo analysis}. The parameters $P$, $M_\star$, $R_\star$, $i$, and $t_i$ are fixed to their value in Table~\ref{tab:final param mcmc}. The limb-darkening coefficients $u_1$ and $u_2$ are fixed for each channel at values interpolated from broadband I, J, H, and K coefficients appropriate for XO-2 \citep{Claret2000}. Among physical parameters, only the planetary radius $R_p$ is a variable; its initial value is that of Table~\ref{tab:final param mcmc}. Similarly to section \ref{sec:Monte Carlo analysis}, the instrumental model is a linear combination of $\alpha^2$, $\alpha$, $x$, $y$, $w$, and $T$, and $F_0$. Each detector column and therefore each spectral channel has its own dependence to instrumental parameters (Figure~\ref{fig:correlations-1param}). The instrumental model is thus computed independently for each channel, as well as for the 6 groups of data:
\begin{equation}
F_{mod}=\Phi(r)\times\big\{F_0\times\Psi(\alpha^2,\alpha,x,y,w,T)\big\}_{g \,\in\,\llbracket 1,6 \rrbracket}
\label{eq:mcmc spec}
\end{equation}

\begin{figure}[htbp]
   \centering
   \includegraphics[width=8cm]{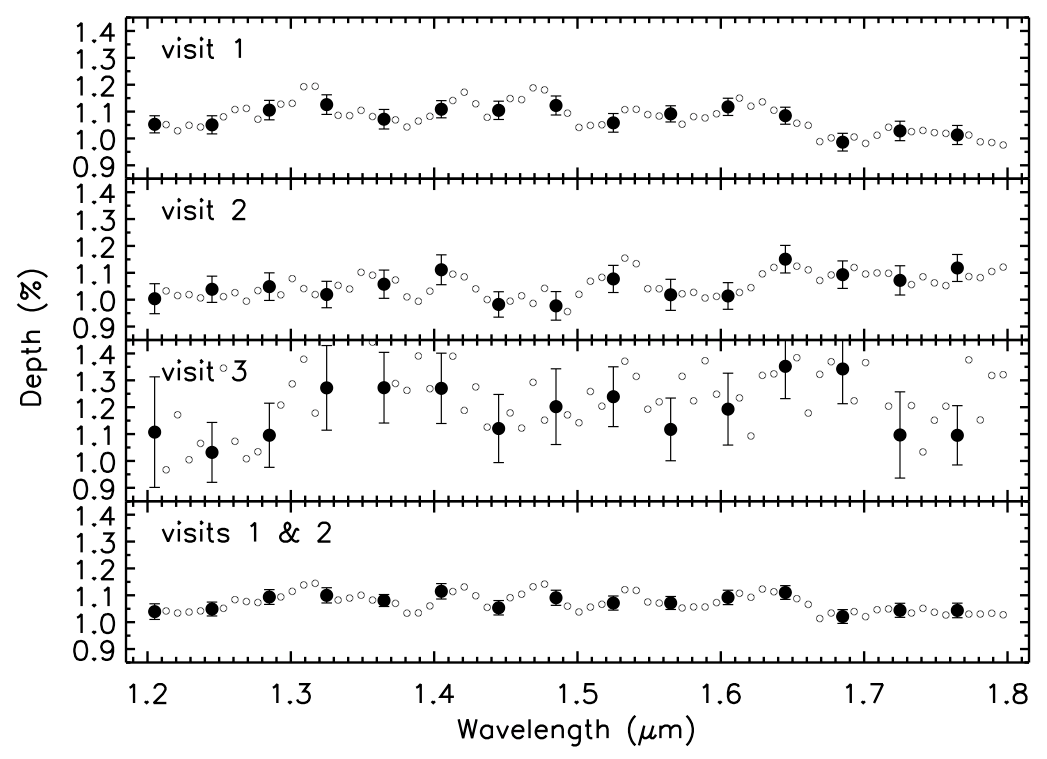}
      \caption{Spectrum of XO-2~N. From top to bottom: 1\ste, 2\nde, 3\rd visit, and joint analysis of the 1\st and 2\nd visits. Independent points are highlighted (filled circles). Intermediate points (open circles) are shown to aid comparison with other NICMOS studies \citep{Swain2009b,Tinetti2010}; their error bars are similar to the highlighted points and are not represented for clarity.}
   \label{fig:spectrum visits}
\end{figure}

A spectrum is built for each visit (Figure~\ref{fig:spectrum visits}). We restrict the range to 1.2-1.8~$\mu$m. In this range, we obtain a standard deviation $\sigma$ of 430, 510, and 1000~ppm for each spectrum for the 1\ste, 2\nde, and 3\rd visit respectively. The approximately twice higher noise for the 3\rd visit than for the other two visits can be explained by larger instrumental parameter variations within orbits, in particular for $y$ and $\alpha$ (Figure~\ref{fig:state parameters}). As a consequence, we ignore the 3\rd visit for the remainder of this analysis. We focus on the 1\st and 2\nd visits. A joint analysis of these 2 visits is performed. We ran the MCMC using the data of both visits, with a single set of transit parameters. Again, the instrumental parameters are optimized for each group. This yields a spectrum with $\sigma=$ 290~ppm (Figure~\ref{fig:spectrum visits}). For comparison, the expected Poisson noise per 5 column channel is 101~ppm for the combination of all observations. The mean absolute difference between the 1\st and 2\nd visit individual spectra is 680~ppm.

\begin{figure}[htbp]
   \centering
   \includegraphics[width=8cm]{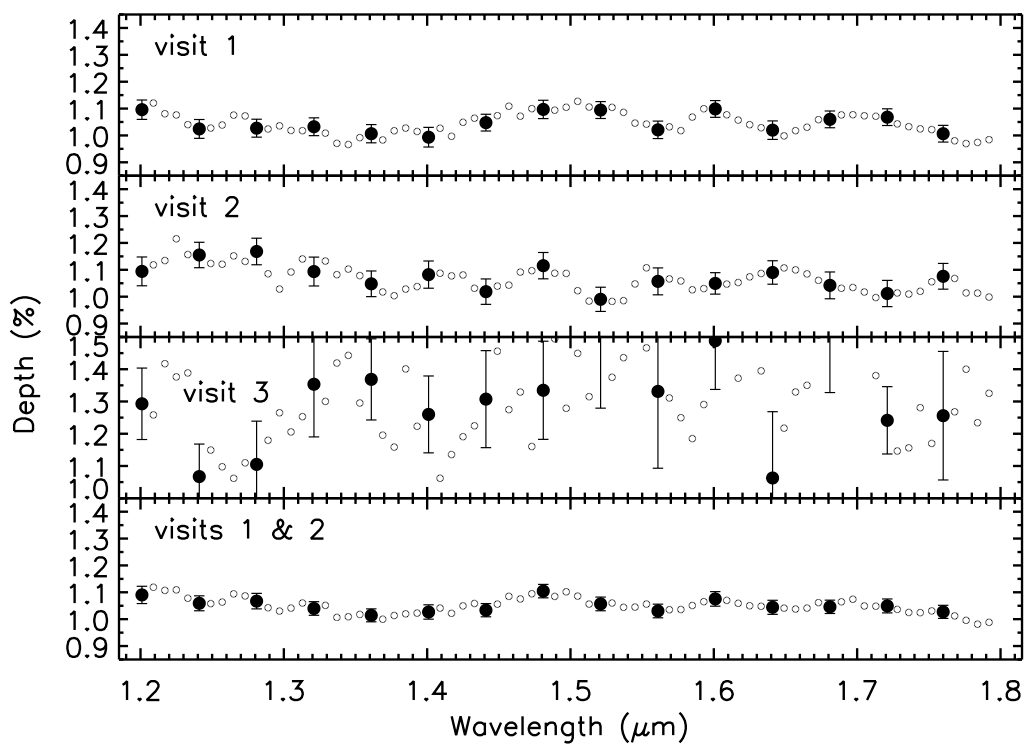}
      \caption{Spectrum of XO-2~S after injection of a flat transit. From top to bottom: 1\ste, 2\nde, 3\rd visit, and joint analysis of the 1\st and 2\nd visits. Independent points are highlighted (filled circles). Intermediate points (open circles) are shown to aid comparison with other NICMOS studies \citep{Swain2009b,Tinetti2010}; their error bars are similar to the highlighted points and are not represented for clarity. The level of systematics is comparable to that of XO-2~N.}
   \label{fig:spectrum visits companion}
\end{figure}

For context, we perform a similar analysis using XO-2~S, a nearly identical star. We extract the channel lightcurves, and numerically inject a simulated transit with a depth constant at all wavelengths. We use the same transit parameters as for XO-2~N. The extracted spectra are shown in Figure~\ref{fig:spectrum visits companion}. Whereas the extracted simulated planetary spectrum should be flat, we find similar variations as we did for XO-2~N: $\sigma=$ 370, 500 and 1480~ppm for the 1\ste, 2\nde, and 3\rd visit respectively, $\sigma=$ 250~ppm for the joint analysis of the 1\st and 2\nd visit, and a mean absolute difference of 610~ppm between the 1\st and 2\nd visit spectra. Therefore, these levels represent the level of remaining NICMOS systematics for these data.

\begin{figure*}[htbp]
   \centering
   \includegraphics[width=12cm]{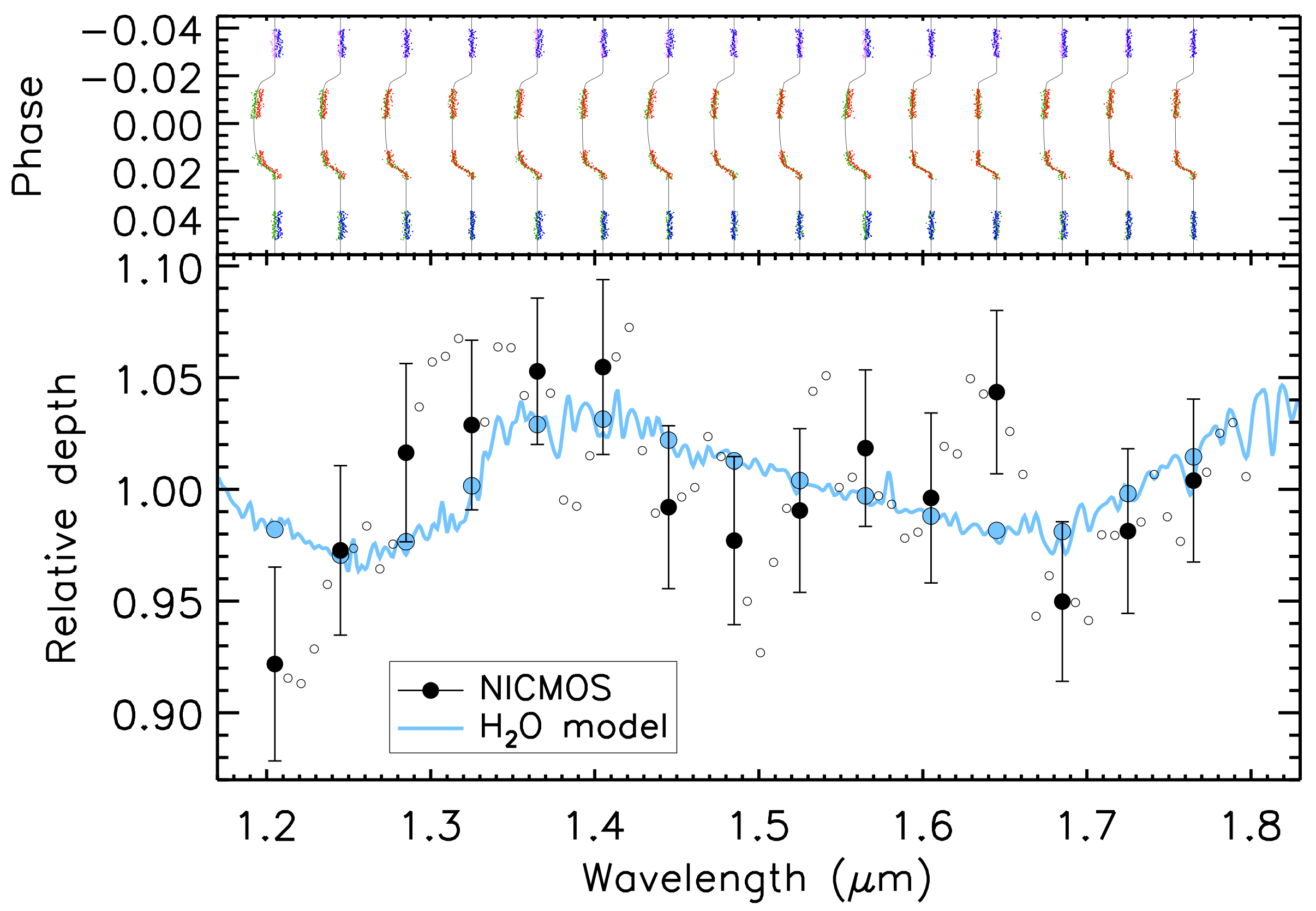}
      \caption{Top: spectroscopic transit lightcurve for XO-2b for the joint analysis of the 1\st and 2\nd visit. Each lightcurve is rotated and vertically offset to the corresponding wavelength. Colors represent groups of data. The best-fit model for each lightcurve is represented as a thin line. Bottom: final spectrum of the planet XO-2b, using the joint analysis of the 1\st and 2\nd visits and XO-2~S as a reference star for XO-2~N. Independent points are highlighted (black filled circles with 1-$\sigma$ error bars). Intermediate points (open circles) are shown to aid comparison with other NICMOS studies \citep[\textit{e.g.}][]{Swain2008,Tinetti2010} ; their error bars are similar to the highlighted points and are not represented for clarity. A synthetic spectrum of an irradiated giant planet including water vapor from \citet{Burrows2010} is shown (blue line), and averaged at the wavelengths of independent data points (blue points).}
   \label{fig:spectrum XO2-b}
\end{figure*}

\begin{figure*}[htbp]
   \centering
   \includegraphics[width=15cm]{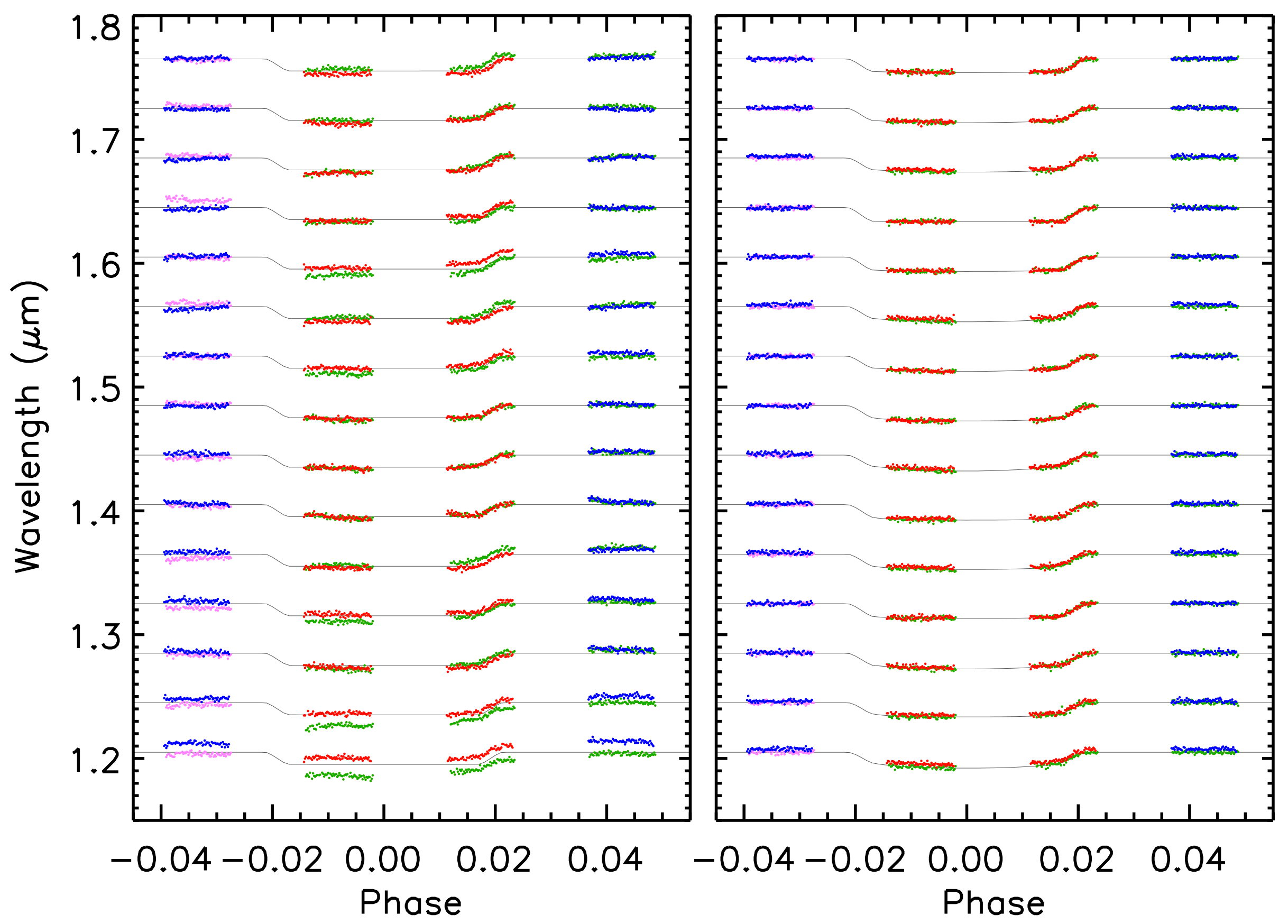}
      \caption{Spectroscopic transit lightcurve for XO-2b for the joint analysis of the 1\st and 2\nd visit, before (left) and after (right) the MCMC decorrelation process. Colors represent groups of data. The best-fit model for each lightcurve is represented as a thin line.}
   \label{fig:lc spectrum XO2-b}
\end{figure*}

Uniquely for XO-2, we can derive a planetary spectrum for XO-2b using XO-2~S as a reference star for XO-2~N. We divide the spectrum of XO-2~N by that of XO-2~S using the joint analysis of the 1\st and 2\nd visits. The result is a normalized spectrum shown in Figure~\ref{fig:spectrum XO2-b}. The corresponding channel lightcurves for XO-2~N before and after the decorrelation process are shown in Figure~\ref{fig:lc spectrum XO2-b}. The corresponding transit depths are reported in Table~\ref{tab:spectrum XO2-b}. The standard deviation of the normalized spectrum is 3.74\%, equivalent to $\sigma=400$~ppm in transit depth. The expected Poisson noise is 142~ppm. Again, using the 1\ste, the 2\nde, or both visits leads to planetary spectra with features that change significantly more than that expected for random noise. Apparently, some instrumental systematics of NICMOS are not common mode and therefore do not cancel. Otherwise, the 1-$\sigma$ uncertainty of the spectrum in Table~\ref{tab:spectrum XO2-b} would have more closely matched the Poisson noise.

\modif{Although systematics are clearly remaining,} we compare \modif{for context} the normalized XO-2b spectrum (obtained by dividing the spectrum of XO-2~N by that of XO-2~S after numerical injection of a flat transit, using the joint analysis of the 1\st and 2\nd visits) to a normalized synthetic spectrum of the irradiated giant planet HD209458b, from \citet{Burrows2010}. In that model, spectral features in the range 1.2-1.8~$\mu$m are largely due to water vapor absorption \citep[see also \textit{e.g.}][for the identification of spectral features]{Brown2001}. In the following, we designate the \citet{Burrows2010} model as water model.

In the range 1.2-1.8~$\mu$m, the mean absolute difference between the data and the water model is 2.70\%, corresponding to 290~ppm in transit depth. The comparison of the data to the water model yields $\chi^2=10.4$ for \modif{$N=15$} degrees of freedom. A comparison of the data to a flat spectrum model (with a constant level adjusted to yield the minimum $\chi^2$) yields $\chi^2=13.5$ for $N=14$ degrees of freedom. \modif{Both models are therefore acceptable fits to the data.} A broad feature is identified around 1.4~$\mu$m, corresponding to an expected water feature. To infer the amplitude of the observed feature, we fit the water model to the data over the range 1.2-1.8~$\mu$m varying the model's amplitude and reference level. The amplitude is calculated as the difference between the minimum and maximum model averaged points, at 1.245 and 1.405~$\mu$m respectively, which corresponds to the amplitude at 1.4~$\mu$m. A $\chi^2$ minimization yields a best-fit amplitude of $580\pm330$~ppm. This level is in good agreement with the theoretical value of 650~ppm albeit with a large uncertainty, and is comparable to those reported for HD 189733b \citep{Swain2008} and XO-1b \citep{Tinetti2010}. \modif{However, this is also comparable to the level of remaining systematics in our data, in particular to the mean absolute difference between the 1\st and 2\nd visit individual spectra. Therefore, this feature may simply be due to systematics.} We do not attempt to fit models with additional molecular species. Interpreted as a 3-$\sigma$ upper limit on the presence of water vapor absorption in the atmosphere of XO-2b, excess absorption at 1.4~$\mu$m is less than 1570~ppm. \modif{Note this simple $\chi^2$ analysis is limited by the fact that residuals may not be white noise.}

\begin{table}[htdp]
\begin{center}
\caption{Planetary spectrum for XO-2b derived from HST NICMOS.}
\begin{tabular}{ccc}
\\
\hline
\hline
$\lambda$ ($\mu$m) & Transit depth (\%) & 1-$\sigma$ uncertainty (\%) \\
\hline

1.205 & 0.978 & 0.046 \\
1.245 & 1.032 & 0.040 \\
1.285 & 1.078 & 0.042 \\
1.325 & 1.091 & 0.040 \\
1.365 & 1.117 & 0.035 \\
1.405 & 1.119 & 0.042 \\
1.445 & 1.052 & 0.039 \\
1.485 & 1.036 & 0.040 \\
1.525 & 1.051 & 0.039 \\
1.565 & 1.080 & 0.037 \\
1.605 & 1.057 & 0.040 \\
1.645 & 1.107 & 0.039 \\
1.685 & 1.007 & 0.038 \\
1.725 & 1.041 & 0.039 \\
1.765 & 1.065 & 0.039 \\

\hline  
\label{tab:spectrum XO2-b}
\end{tabular}
\end{center}
\end{table}

\section{Discussion}
\label{sec:Discussion}

\subsection{XO-2b}

Models predict water vapor in the atmosphere of hot Jupiters as a major absorber in the near-infrared \citep{Fortney2008,Fortney2010,Burrows2010}. In this domain, pM and pL class planets exhibit the same spectral features, with a larger amplitude for the pM class due to a greater atmospheric scale height. \citet{Fortney2010} calculated models of transmission spectrum as a function of the planet temperature for two surface gravities. With surface gravity $g=14.71\rm\;m\,s^{-2}$ \citep{Fernandez2009} and dayside temperature $T=1500$ K \citep{Machalek2009}, these models predict a water vapor absorption feature at 1.4~$\mu$m that would increase the apparent planetary radius by 3\%, \textit{i.e.} the transit depth by 600~ppm. A similar feature is predicted by \citet{Burrows2010}, which we show in Figure~\ref{fig:spectrum XO2-b}. Our XO-2b spectrum shows an increase in absorption of $580\pm330$~ppm, in good agreement with these predictions, albeit with a large uncertainty. \modif{Given the level of remaining systematics in our data, this feature may however be spurious}. Our 3-$\sigma$ upper limit on the 1.4~$\mu$m molecular absorption, 1570~ppm, is consistent with these models.

\subsection{NICMOS}

Transit spectroscopy is at the limit of NICMOS capabilities. With the simplest data analysis, i.e. without a decorrelation process or alternative approach, the derived planetary spectrum of XO-2b (Figure 8) exhibits noise-like features that are approximately one order of magnitude greater than signals anticipated from a planetary atmosphere. The dominant noise source is due to the combination of the different positioning of the spectral trace on the detector, notably the angle $\alpha$ of the spectral trace, combined with intra-pixel variations. Previous NICMOS measurements presumably are affected similarly as well.
 
In this work, we decorrelated the NICMOS data for XO-2b with the MCMC technique using a quadratic function for the spectral tilt angle $\alpha$ and linear functions for the other parameters. Previous studies (e.g. \citet{Tinetti2010}) have performed similar parametric analyses, and others have used alternative methods such as Gaussian processes (e.g. \citet{Gibson2012a,Gibson2012b}). For the NICMOS data of XO-2b, the availability of a nearly-identical companion star (XO-2~S) observed simultaneously provides a unique opportunity for data analysis and calibration. The presence of this bright, nearly-identical reference star could be used by other investigators as a benchmark to test their NICMOS lightcurve and spectral extraction methods. 

To use the companion star XO-2~S in our analysis, we numerically injected into the data of XO-2~S a transit with a depth constant in wavelength, and then derived a spectrum with the same procedure as for the planet host star, XO-2~N. The derived simulated spectrum (of XO-2~S) has residuals comparable to those evident in the planetary spectrum derived for XO-2~N, but which can only be due to instrumental effects or effects of the analysis because we injected a flat, featureless spectrum. In addition, the residuals exhibited in the synthetic spectrum (of XO-2~S) differ for each visit: they have a typical scale of 0.05~$\mu$m in wavelength and 500~ppm in amplitude. Therefore, features of that scale derived by our technique cannnot be confidently ascribed to molecular signatures in XO-2b. We note that the spectrum derived for XO-1b in the Appendix exhibits similar residuals, which also differ for the two visits. In both cases, XO-2b and XO-1b, \modif{a broader 0.2~$\mu$m-wide feature appears at 1.4~$\mu$m, where extant models predict extra absorption due to water vapor. However, the observed feature has an amplitude comparable to the level of systematics and is thus of weak statistical significance.}

Our MCMC decorrelation process may not correct for all systematics. In particular, \citet{Pont2009} and \citet{Gibson2011a} suggest that the in-transit instrumental parameters should be completely sampled during the out-of-transit orbits. This is unfortunately not the case for the XO-2 data, in particular for the angle of the spectral trace. Whereas the need for sophisticated analysis techniques is clear, different analysis of previous NICMOS data lead to different results \citep{Swain2008,Tinetti2010,Gibson2011a}. This supports our remark that extracting planetary spectra is at the limit of NICMOS capabilities. 

Whether the above conclusions apply only to those data that we analyzed (XO-1 and XO-2), and/or only for our analysis technique, or applies more generally to all NICMOS spectra remains an open question, one that is beyond the scope of this analysis. A program to analyze all relevant NICMOS data has been approved under HST archival program 12844, "Exoplanetary Spectroscopy with  NICMOS Revisited'' (Deming, P.I.). 

Planetary spectra have been observed with HST's new Wide Field Camera 3 (WFC3), which has much smaller, indeed as yet unmeasurable, intra-pixel sensitivity variations than NICMOS, for which they are an important effect \citep{Pavlovsky2011}. Using WFC3 \citet{Berta2012} reported a result for the planet orbiting the H = 9.1 magnitude M dwarf GJ1214, and showed that the results are nearly Poisson-noise limited after a few corrections are carefully applied to the WFC3 data.

~\\
We are grateful to Jeff Valenti, Ronald L. Gilliland, Christopher M. Johns-Krull, and Ed Nelan for their participation on the observation proposal, and for a careful reading of the paper. We thank Adam Burrows for providing us with a hot Jupiter spectrum model. We thank an anonymous referee for comments that significantly improved the manuscript. Financial support for this work was provided through program HST GO-11228 from STScI and the NASA Origins of Solar Systems grant NNX10AG30G.

\appendix

\section{Analysis of XO-1 with NICMOS}
\label{sec:Analysis of XO-1 with NICMOS}

NICMOS spectroscopic observations of XO-1b have been analyzed previously by two teams. \citet{Tinetti2010} identify molecular components in the atmosphere of the planet. From the same data, \citet{Gibson2011a} find no conclusive evidence for such features, and attribute residuals in the spectrum to NICMOS instrumental noise. Here, we briefly present another re-analysis of these same data.

XO-1b was observed during 2 transit events, on 2008 February 10 and 2008 February 21. Each visit consists of 5 HST orbits. A total of 566 images were acquired, each with an exposure time of 40 s. The 1\st orbit of each visit is rejected. In addition, the first 10 images of each orbit are affected by strong FWHM and temperature variations. This represents a small fraction of the data (18\%), and these images are rejected. Unlike XO-2, no comparison star is present in the field of view. 

\begin{figure}[htdp]
   \centering
   \includegraphics[width=10cm]{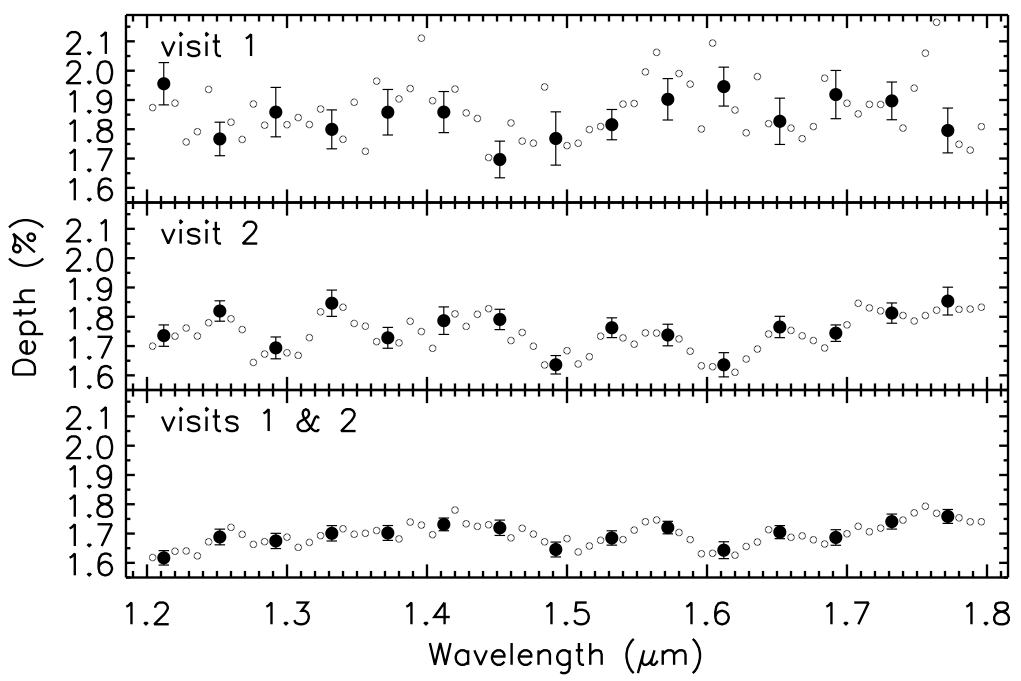}
      \caption{Spectrum of the planet XO-1b, using the 1\st (top), 2\nd (middle), and both visits simultaneously (bottom). Independent points are highlighted (black filled circles with 1-$\sigma$ error bars). Intermediate points (open circles) are shown to aid comparison with other NICMOS studies \citep{Swain2009b,Tinetti2010}; their error bars are similar to the highlighted points and are not represented for clarity.}
   \label{fig:spectrum XO-1}
\end{figure}

The analysis procedure is equivalent to that of XO-2 as described in the body of this paper. We examine which parameters to include via trial runs. In the case of XO-1, we use the same first-order parameters, and also include an $x^2$ term because it yields a lower dispersion than $\alpha^2$. We thus construct the spectrum using the MCMC technique with a decorrelation of the instrumental parameters $F_0$, $x$, $y$, $\alpha$, $w$, and $T$, and $x^2$. We construct a planetary spectrum from the 1\st and 2\nd visits independently, and for both visits analyzed simultaneously (Figure~\ref{fig:spectrum XO-1}). We obtain a standard deviation of 1020, 710 and 430~ppm respectively. The mean absolute difference between the spectra obtained from the 1\st and 2\nd visit is 1590~ppm. We consider the spectra obtained from both visits simultaneously as our final spectrum (Table~\ref{tab:spectrum XO1-b}). The corresponding channel lightcurves before and after the decorrelation process are shown in Figure~\ref{fig:lc spectrum XO1-b}.

\begin{figure}[htdp]
   \centering
   \includegraphics[width=12cm]{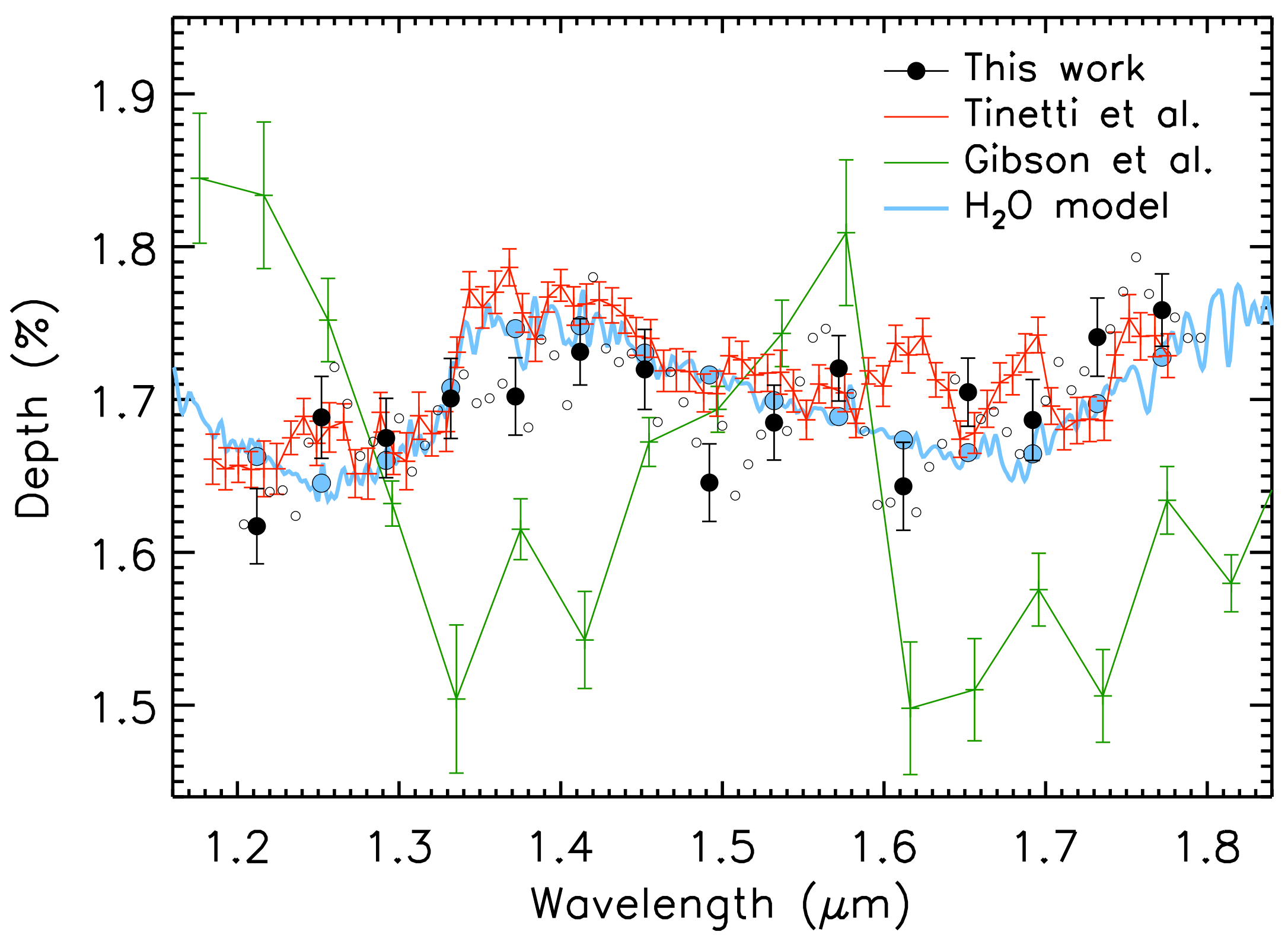}
      \caption{Spectrum of XO-1b derived from this work using the joint analysis of the 1\st and 2\nd visits (black), from \citet{Tinetti2010} using the 2\nd visit only (red), and from \citet{Gibson2011a} using the 2\nd visit only (green). In our spectrum, independent points are highlighted (black filled circles with 1-$\sigma$ error bars). Intermediate points (open circles) are shown to aid comparison; their error bars are similar to the highlighted points and are not represented for clarity. A synthetic spectrum of an irradiated giant planet including water vapor from \citet{Burrows2010} is shown (blue line), and averaged at the wavelengths of independent data points (blue points).}
   \label{fig:spectrum XO-1 superp}
\end{figure}

\begin{figure}[htdp]
   \centering
   \includegraphics[width=15cm]{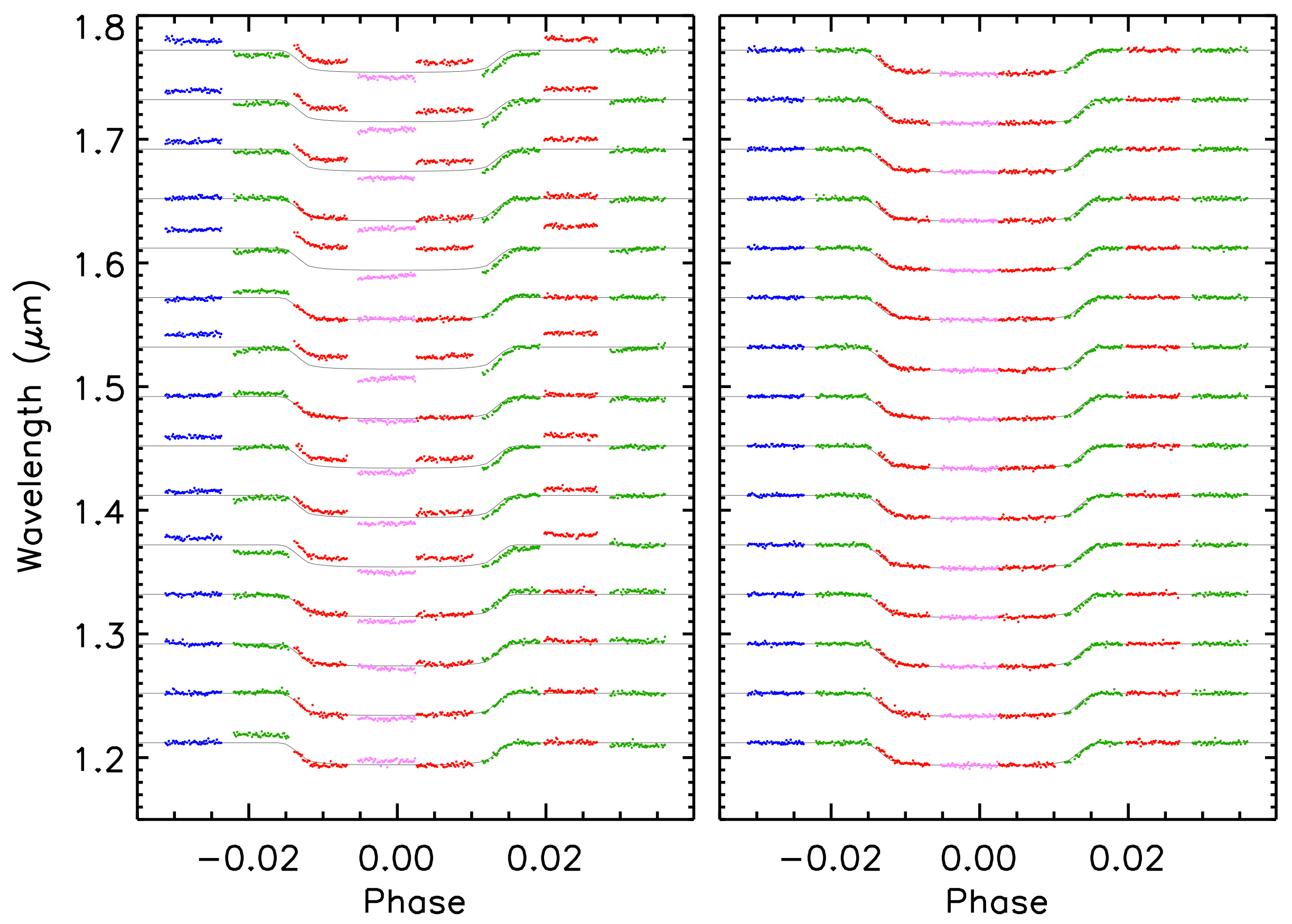}
      \caption{Spectroscopic transit lightcurve for XO-1b derived from this work using the joint analysis of the 1\st and 2\nd visits, before (left) and after (right) the MCMC decorrelation process. Colors represent groups of data. The best-fit model for each lightcurve is represented as a thin line.}
   \label{fig:lc spectrum XO1-b}
\end{figure}

We compare our spectrum to previous results \citep{Tinetti2010,Gibson2011a}. The mean absolute difference is 370~ppm between our spectrum and that of \citet{Tinetti2010}, and 1230~ppm between our spectrum and that of \citet{Gibson2011a} (Figure~\ref{fig:spectrum XO-1 superp}). \modif{However, both \citet{Tinetti2010} and \citet{Gibson2011a} used the second visit only. Using our spectrum obtained from the second visit, the mean absolute difference is 730~ppm between our spectrum and that of \citet{Tinetti2010}, and 1520~ppm between our spectrum and that of \citet{Gibson2011a}. Our final spectrum globally differs to that of \citet{Gibson2011a}, and differs significantly to that of \citet{Tinetti2010} in particular at scales attributed to specific molecular features of CH$_4$, CO$_2$, and CO.}

We attribute the large difference between the \citet{Gibson2011a} spectrum of XO-1b and that of \citet{Tinetti2010}, and this work, with the decorrelation from instrumental parameters such as the spectrum's tilt and position in $x$ and $y$ that the latter two analyses included but \citet{Gibson2011a} did not. \citet{Gibson2011a} deliberately excluded these parameters to avoid extrapolations. We included those parameters because they correspond to physically motivated instrumental effects and were necessary in the case of XO-2. In particular, as shown in this work, and also described by \citet{Deroo2010}, the spectrum tilt is a significant contributor to systematics in the derived spectrum from NICMOS data. The differences between the this work's spectrum of XO-1b and that of \citet{Tinetti2010} are smaller, but reinforce our conclusion that NICMOS spectra are marginal for detecting molecular features in the atmospheres of gas giant extrasolar planets.

For context, we also compare our XO-1b spectrum obtained using the joint analysis of the 1\st and 2\nd visits to the same theoretical spectrum as in Section \ref{sec:Spectrum Monte Carlo analysis}, which we designate as a water model. After scaling to the XO-1 parameters, this model has a mean of 1.802\% in the range 1.2-1.8~$\mu$m, 1.06 times larger than the mean of the data. We rescale the model by that factor. In the range 1.2-1.8~$\mu$m, the mean absolute difference between our data and the water model is 310~ppm. The comparison of our data to the water model yields $\chi^2=30$ for \modif{$N=15$} degrees of freedom. A comparison of the data to a flat spectrum model (with a constant level adjusted to yield the minimum $\chi^2$) yields $\chi^2=33.6$ for $N=14$ degrees of freedom. \modif{Both models are therefore acceptable fits to the data.} A broad feature is identified around 1.4~$\mu$m, corresponding to an expected water feature. To infer the amplitude of the observed feature, we follow the same procedure as in Section \ref{sec:Spectrum Monte Carlo analysis}. We find a best-fit amplitude of $590\pm200$~ppm, about 2 times lower than the theoretical value (1050~ppm). Again, this level is comparable to those reported for HD 189733b \citep{Swain2008} and XO-1b \citep{Tinetti2010}. \modif{However, given the discrepancy between the 1\st and 2\nd visit individual spectra, this feature may simply be due to systematics.} We do not attempt to fit models with additional molecular species. Interpreted as a 3-$\sigma$ upper limit on the presence of water vapor absorption in the atmosphere of XO-1b, excess absorption at 1.4~$\mu$m is less than 1190~ppm.

\begin{table}[htdp]
\begin{center}
\caption{Planetary spectrum for XO-1b derived from HST NICMOS.}
\begin{tabular}{ccc}
\\
\hline
\hline
$\lambda$ ($\mu$m) & Transit depth (\%) & 1-$\sigma$ uncertainty (\%) \\
\hline
1.212  &   1.617  &   0.025 \\
1.252  &   1.688  &   0.027 \\
1.292  &   1.675  &   0.026 \\
1.332  &   1.701  &   0.026 \\
1.372  &   1.702  &   0.025 \\
1.412  &   1.731  &   0.022 \\
1.452  &   1.720  &   0.026 \\
1.492  &   1.646  &   0.025 \\
1.532  &   1.685  &   0.025 \\
1.572  &   1.720  &   0.021 \\
1.612  &   1.643  &   0.029 \\
1.652  &   1.705  &   0.022 \\
1.692  &   1.687  &   0.027 \\
1.732  &   1.741  &   0.026 \\
1.772  &   1.759  &   0.024 \\
\hline  
\label{tab:spectrum XO1-b}
\end{tabular}
\end{center}
\end{table}

\FloatBarrier

\bibliography{biblio.bib}

\end{document}